\documentclass[a4paper,10pt,twocolumn]{article}
\pdfoutput=1
\usepackage[a4paper, total={18cm, 26cm}]{geometry}
\usepackage{times} 

\usepackage[binary-units,per-mode=symbol,detect-all,forbid-literal-units]{siunitx}
\usepackage{graphicx}
\usepackage{amsfonts}
\usepackage{amsmath}
\usepackage{amsthm}
\usepackage{url}
\usepackage[american]{babel}
\usepackage{booktabs}
\usepackage{multirow}
\usepackage{float}
\usepackage{enumitem}
\usepackage{xspace}
\usepackage[sort&compress,square,numbers]{natbib}

\usepackage[caption=false]{subfig}  % sottofigure, sottotabelle
\usepackage[skip=2pt,font=footnotesize,labelfont=sf,textfont=sf]{caption}

\usepackage[capitalise]{cleveref}
\usepackage{tcolorbox}

\newtcolorbox[auto counter,number within=section,
crefname={Proposition}{Propositions}]%
{myproposition}[2][]{colback=white,colframe=gray!25!black,fonttitle=\bfseries,
title=Prop. \thetcbcounter: #2,#1}

\newtcolorbox[auto counter,number within=section,
crefname={Definition}{Definitions}]%
{mydefinition}[2][]{colback=white,colframe=gray!25!black,fonttitle=\bfseries,
title=Def. \thetcbcounter: #2,#1}

\newtcolorbox[auto counter,number within=section,
crefname={Highlight}{Highlights}]%
{highlightBox}[2][]{colback=white,colframe=gray!25!black,fonttitle=\bfseries,
title=Highlight \thetcbcounter: #2,#1}

\newtheoremstyle{defStyle}
  {\topsep}   % ABOVESPACE
  {\topsep}   % BELOWSPACE
  {\normalfont}  % BODYFONT
  {2pt}       % INDENT (empty value is the same as 0pt)
  {\itshape} % HEADFONT
  {}         % HEADPUNCT
  {5pt plus 1pt minus 1pt} % HEADSPACE
  {\thmname{#1} \thmnumber{#2}: \thmnote{\bfseries#3}}          % CUSTOM-HEAD-SPEC

\theoremstyle{defStyle}
\newtheorem{definition}{Def.}[section]

\crefname{figure}{Figure}{Figures}

\usepackage[nohyperlinks]{acronym}
\usepackage[yyyymmdd]{datetime}

\acrodef{BC}{Blockchain}
\acrodef{CHF}{Cryptographic Hash Function}
\acrodef{DB}{Data Base}
\acrodef{DBMS}{Database Management System}
\acrodef{DDoS}{Distributed Denial of Service}
\acrodef{DLT}{Distributed Ledger Technology}
\acrodefplural{DLT}{Distributed Ledger Technologies}
\acrodef{DPoS}{Delegated Proof of Stake}
\acrodef{IoT}{Internet of Things}
\acrodef{PBFT}{Practical Byzantine Fault Tolerance}
\acrodef{PKI}{Public Key Infrastructure}
\acrodef{PoW}{Proof of Work}
\acrodef{PoS}{Proof of Stake}
\acrodef{PoN}{Proof of Networking}
\acrodef{TPS}{Transactions per Second}

\newcommand{\blockprop}{\ensuremath{\mathit{B_{P}}}\xspace}
\newcommand{\blockinterval}{\ensuremath{\mathit{B_{GI}}}\xspace}

\newcommand{\miningOverhead}{\ensuremath{\mathit{AMO}(R, \blockprop, \blockinterval)}\xspace}

\newcommand{\terahash}{\ensuremath{\si{\tera}\text{H}}\xspace}
\newcommand{\Eth}{Ethereum\xspace}

\newtheorem{Pitf}{Pitfall}[section]

\begin{document}
\sloppy
\title{\vspace{-0.5cm}\bfseries What is a Blockchain? A Definition to Clarify the Role\\  of the Blockchain in the Internet of Things}

\author{Lorenzo Ghiro,$^1$ Francesco Restuccia,$^2$ Salvatore D'Oro,$^2$ Stefano Basagni,$^2$ \\
Tommaso Melodia,$^2$  Leonardo Maccari,$^3$ Renato Lo Cigno$^4$ \\
\textit{$^1$University of Trento, Italy,
$^2$Northeastern University, USA,}\\
\textit{$^3$University of Venice, Italy,
$^4$University of Brescia, Italy}
}

\date{~}

\maketitle

\begin{abstract} 
The use of the term \emph{blockchain} is documented for disparate projects,
from cryptocurrencies to applications for the Internet of Things (IoT), and many more.
The concept of blockchain appears therefore blurred, as it is hard to believe that the same technology can empower applications that have extremely different requirements and exhibit dissimilar performance and security.
This position paper elaborates on the theory of distributed systems to advance a clear definition of blockchain that allows us to clarify its role in the IoT. 
This definition inextricably binds together three elements that, as a whole, provide the blockchain with those unique features that distinguish it from other distributed ledger technologies: \emph{immutability}, \emph{transparency} and \emph{anonimity}.
We note however that immutability comes at the expense of remarkable resource consumption, transparency demands no confidentiality and anonymity prevents user identification and registration.
This is in stark contrast to the requirements of most IoT applications that are made up of resource constrained devices, whose data need to be kept confidential and users to be clearly known.
Building on the proposed definition, we derive new guidelines for selecting the proper distributed ledger technology depending on application requirements and trust models, identifying common pitfalls leading to improper applications of the blockchain.
We finally indicate a feasible role of the blockchain for the IoT: myriads of local, IoT transactions can be aggregated off-chain and then be successfully recorded on an external blockchain as a means of public accountability when required.
\end{abstract}

\acresetall

\section{Introduction}
\label{sec:intro}

The \emph{blockchain} came into the limelight with the advent of the Bitcoin cryptocurrency,
by far the most successful blockchain application, which in January of 2021 set its new
record of market capitalization exceeding 758 billions of US dollars.
The blockchain features observed in Bitcoin, i.e., 
decentralization, resistance to powerful cyberattacks and preservation of users privacy,
raised the enthusiasm of many research communities. 
This enthusiasm led to an extremely large number of disparate proposals
for using the blockchain in many different applications, including
Supply Chain Management~\cite{montecchi2019s,tse2017,tian2016agri,miller2018},
E-Voting \cite{bcEvoting,ayed2017,bistarelli2017,liu2017voting,hardwick2018voting,wang2018large,qi2017,noizat2015},
Smart Grid \cite{Gai2019,lombardi2018,mylrea2017,Mengelkamp2017,munsing2017,hahn2017smart,laszka2017,yan2017},
Healthcare \cite{Xu2019,dwivedi2019,bcHealthcare,Yue2016}, Banking \cite{Cocco2017,Guo2016},
Smart Cities \cite{Shen2019,ibba2017city,patil2017}, and even Vehicular and Aerial Networks \cite{Liu2019,bcVehicular,Yang2019,patel2019vehichain,cebe2018,
michelin2018,zhang2018vanet,li2018coin4Vehic,rowan2017,sharma2017,
singh2017,yang2017,yuan2016,leiding2016,kapitonov2017,liang2017,bcDrones,ferrer2016,bcUAV2}.
Surveys abound on the efforts of applying the blockchain also to the many expressions of the Internet of Things (IoT)~\cite{Wu2019,Viriyasitavat2019,wang2019survey,xu2017taxonomy,panarello2018blockchain,zheng2018blockchain,lin2017survey,lin2017surveyNetSec,novo2018blockchain}. 
This vast application range makes the blockchain seem a \emph{universal} technology, no longer limited only to cryptocurrencies but also capable to empower most IoT applications, addressing their multiple vulnerabilities~\cite{IoTsecSurvey}.

This apparent universality of the blockchain looks suspicious, suggesting that the term is used with many different meanings. 
Considering that the Bitcoin blockchain currently supports the validation of less than 10 \ac{TPS} and exhibits a power consumption similar to that of an industrialized country such as Ireland~\cite{deVries2018}, it is dubious that it can support the millions of IoT TPS~\cite{kim2019advanced} and meet the typical IoT power constraints.
In fact, we notice that moving to application domains different from cryptocurrencies,
the original characteristics of the Bitcoin blockchain have been diluted, if not completely transformed, leading to possible misunderstanding and confusion. 
On the one hand, we have the \emph{permissionless} blockchains like Bitcoin~\cite{vademecum}, celebrated for their \ac{PoW}-based cryptographical security, their decentralization and strict privacy defense through anonymity.
On the other hand, many proposed ``blockchains'' are \emph{permissioned}, requiring user identities to be registered with some trusted authority (or consortium) and whose internal security does not depend on some hard cryptographical problem like the \ac{PoW}.
As a consequence, the term blockchain appears to be overloaded, and therefore ambiguous, as it is used to indicate 
ledger technologies that under the hood obtains security, performance and decentralization in completely different ways.

This position paper analyzes the multiple technologies proffered under the term \emph{blockchain} and
proposes a clear definition of blockchain that allows us to argue about its role in the IoT.
Building on the theory of distributed systems and on the critical analysis of current
blockchain applications, the definition identifies three elements that, only when combined together,
give to blockchains their specific features of openness, decentralization, security and ability to preserve user privacy.
These three elements are: 
relying on a \textsc{Strong Distributed Consensus Protocol}, which makes the blockchain immutable, hence secure from tampering attacks, and further frees the system from centralized trusted authorities (e.g., banks);
maintaining a \textsc{Full \& Public History of Transactions}, which permits their distributed and completely transparent validation, and 
being \textsc{Open to Anonymous Users}, thus allowing blockchains to preserve users privacy.
  
A definition for blockchain based on these three elements results quite restrictive, amplifying the voice of those who also advocate a strict definition of blockchain~\cite{Ammous2016,Garzik2018,perez2016blockchain}.
However, clarifying the nature of the blockchain with this clear definition allows us to argue that the blockchain  is not the universal and limitless technology that may seem to be.
In fact, we state that blockchains are beneficial only for a limited range of applications, and that their integration into the IoT domain is not appropriate.
We substantiate these arguments drawing new guidelines for the proper adoption of the blockchain, suggesting clearly when the blockchain should \emph{not} be adopted.
We do so highlighting a list of common scenarios where the applications requirements conflict
with the blockchain characteristics. 
For example, the use of a blockchain for storing sensitive information is a pitfall,
because the blockchain immutability would prevent the compliance with regulations that demand user data to be erasable upon request.

The remainder of this paper includes a review of the fundamental principles of the blockchain (\cref{sec:background})
and of its theoretical roots, namely, distributed consensus protocols (\cref{sec:distr-cons}).
We acknowledge that the analysis proposed in this paper is partial, meaning that the goal of the paper is to \emph{restrict} the meaning of blockchain, rather than building an exhaustive list of all its possible meanings and uses, as a standard survey would do. 
Still, we provide abundant academic references in support 
of our definition and highlight that there are only benefits ---for the scientific community, industry and practitioners at large---
to restrict and disambiguate the use of this term and rather using different ones, inventing them if necessary,
to identify technologies and solutions that are far away from the original use of the term. 
In \cref{sec:demystify-blockchain} we condense the specific features of blockchains into a connotative definition,
based on which we build the guidelines for the proper adoption of the blockchain technology. 
We then discuss popular applications that contrast with the proposed definition of blockchain in~\cref{sec:pitfalls}, 
explaining which application requirements clash with the blockchain features.
In \cref{sec:roleBCwithIoT} we indicate a possible role for blockchains in the IoT, namely,
they can play as complementary (external) ledger services.
Final remarks are drawn in \cref{sec:conclusion}.

\section{Blockchain Fundamentals}\label{sec:background}

\begin{figure}[htb]
  \centering
    \includegraphics[width=\linewidth]{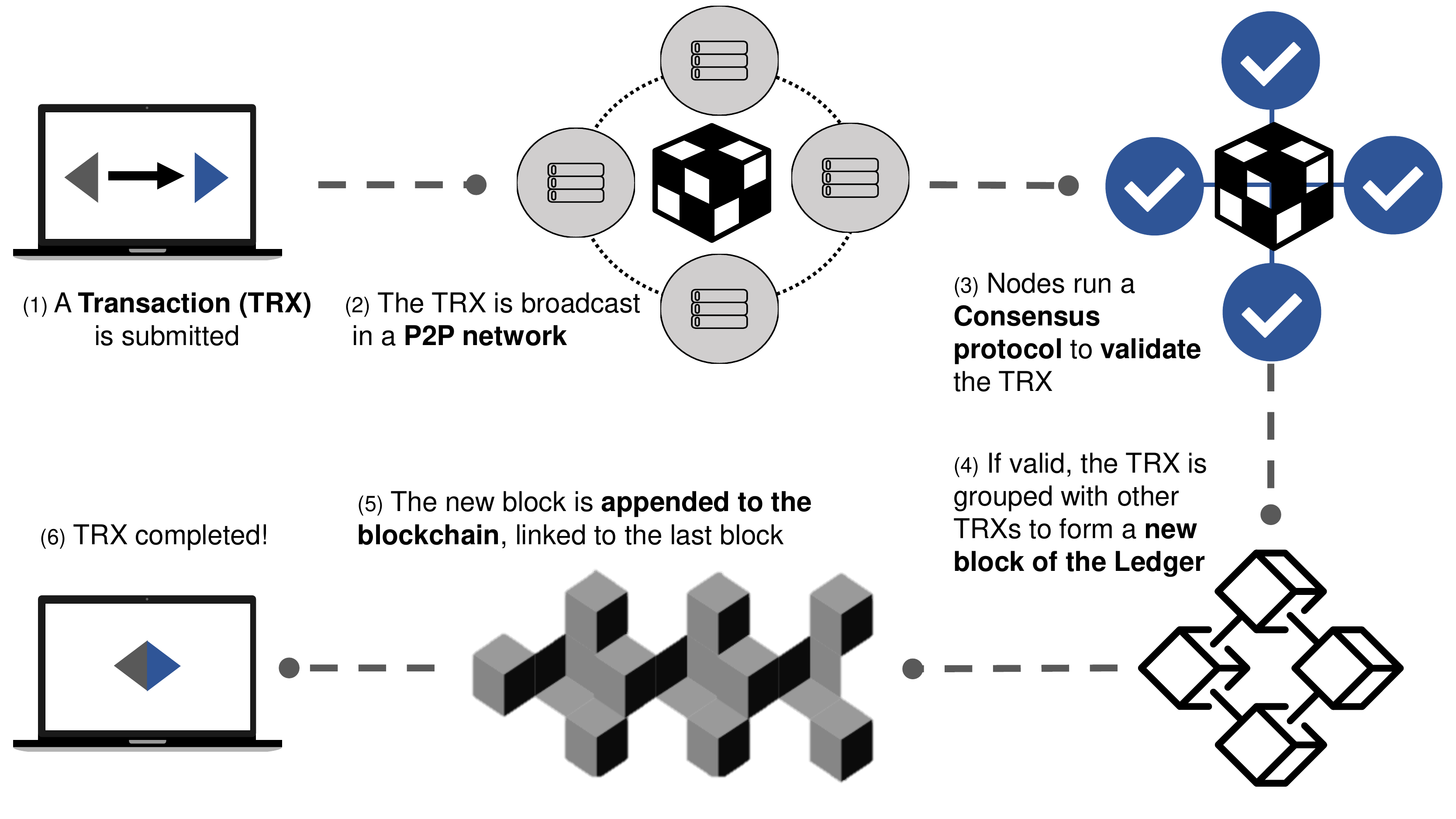} 
    \caption{Processing of a transaction before storage in the blockchain.}
\label{fig:TXworkflow}
\end{figure}

\Cref{fig:TXworkflow} is proposed to gently introduce the blockchain fundamentals starting
from an illustration of the life-cycle of a transaction, which will be finally registered in a blockchain.
Everything starts when a transaction is issued, e.g., because a smart device is querying a remote service and pays to access the data. 
The transaction is announced in the P2P network and received by validator nodes. 
These nodes run a consensus protocol to decide about the validity of the transaction. 
If they reach a consensus on the fact that the device really owns the resources that is about to spend,
then the transaction is considered valid.
If so, it is grouped with others recently approved, forming a new block of transactions
that will be registered in the ledger by appending it to the blockchain. 
At the end, the success of the transaction is notified to the users and the data is transferred to the device.

At first glance, the blockchain can be defined as the plain data structure used to record transactions. 
However, from a broader perspective, a blockchain can be considered a distributed system that in general includes: 

\begin{itemize}[itemsep=0em, leftmargin=0.5cm]

  \item A \textit{Peer-to-Peer (P2P) network} made of all those nodes that either read or cooperatively write transactions in the blockchain, and
  
  \item a \textit{consensus protocol}, namely, a set of policies agreed upon and implemented by all nodes, which are the rules that regulate which and how new transactions can be added to the blockchain.

\end{itemize}

A blockchain turns out to be a possible implementation of a Shared Ledger.
The group of entities allowed to write new transactions in the Shared Ledger, appending them to the blockchain,
can vary from few selected and authenticated users up to any anonymous user. 
These different writing privileges depend on the rules of the chosen consensus protocol,
which are decisive to determine if the resulting Shared Ledger will be \textit{public} or \textit{private}.
    
\paragraph{ Public or Permissionless Ledger}
In a public (permissionless) ledger, the record of transactions is public and the consensus protocol is open to anybody.
This means that 
i) anyone in the world can verify the correctness of the ledger and
ii) even anonymous strangers without explicit permission can join the network and participate
in the validation process of transactions, provided only that they comply with the consensus protocol. 
The absence of any form of control on users, that are not accountable as they are anonymous,
is a an issue for the security of the ledger.
To counter this lack of trust and still ensure security, the usual consensus protocol of a permissionless ledger imposes
stringent conditions to be met upon proposing a new block of transactions.
Such conditions are so severe that, somehow, prove the honest commitment of the proposer.
For example, in both Bitcoin and \Eth --- the most iconic permissionless blockchains---
the proposer of a new block of transactions must provide the so-called \acf{PoW}, 
which is the solution to a very hard cryptographical problem. 
This mechanism secures the ledger discouraging malicious users, but hampers the ledger performance as well.
Consider, for example, that the number of \ac{TPS} processed by Bitcoin and \Eth is on average below 20 TPS,
a very limited throughput if compared to the tens of thousands processed by Visa~\cite{vujivcic2018blockchain}.
Moreover, in the Visa platform transactions are recorded sequentially, not in blocks.
For this reason the transaction latency, i.e., 
the interval of time between submission and recording of a transaction,
can be kept down to at most a few seconds with Visa,
while in Bitcoin the same latency averages tens of minutes and can grow up to several hours. 

\paragraph{ Private or Permissioned Ledger}
Private ledgers arose as an attempt to improve performance and to have more control on users. 
These ledgers are typically implemented by big corporations or banks, so to have a common platform to share business information among few and well known partners. 
A shared and mutual level of trust can be given for granted, as only registered (hence accountable) entities have
the permission to write data into the blockchain.
The security of permissioned blockchains depends therefore on classical authentication mechanisms
rather than on the mathematical strength of techniques such as the PoW. 
The trust model resulting from permissions allows blockchain managers
to replace the resource-hungry consensus protocols of permissionless blockchains
with more traditional, efficient, and faster ones. 
Such faster consensus protocols are necessary to support critical business operations,
whose recording cannot tolerate the typical low transaction rates and high inefficiencies of permissionless ledgers.

For the first time, permissionless ledgers free users from trusted authorities, such as a \ac{PKI} or banks,
with the added novelty that the validation process happens transparently in public, without uncovering the identity of users.
The trust required to maintain a public, permissionless ledger open to anonymous users is given by the consensus protocol only.
A permissionless blockchain can thus be considered as a \textit{trust builder in a trustless network} and the enabler of
an \textit{open, privacy-preserving, disintermediated marketplace}.

\subsection{The Need of the Transactions History}\label{needTRXhistory}

Validators need the history of transactions to determine who and how many resources each user owns,
an indispensable knowledge to validate new transactions. 
However, building this history in a distributed system is complicated by the double spending problem, briefly described below.

\subsubsection*{\textbf{Double Spending Problem}} 
\begin{figure}[h]
  \centering
    \includegraphics[width=\linewidth]{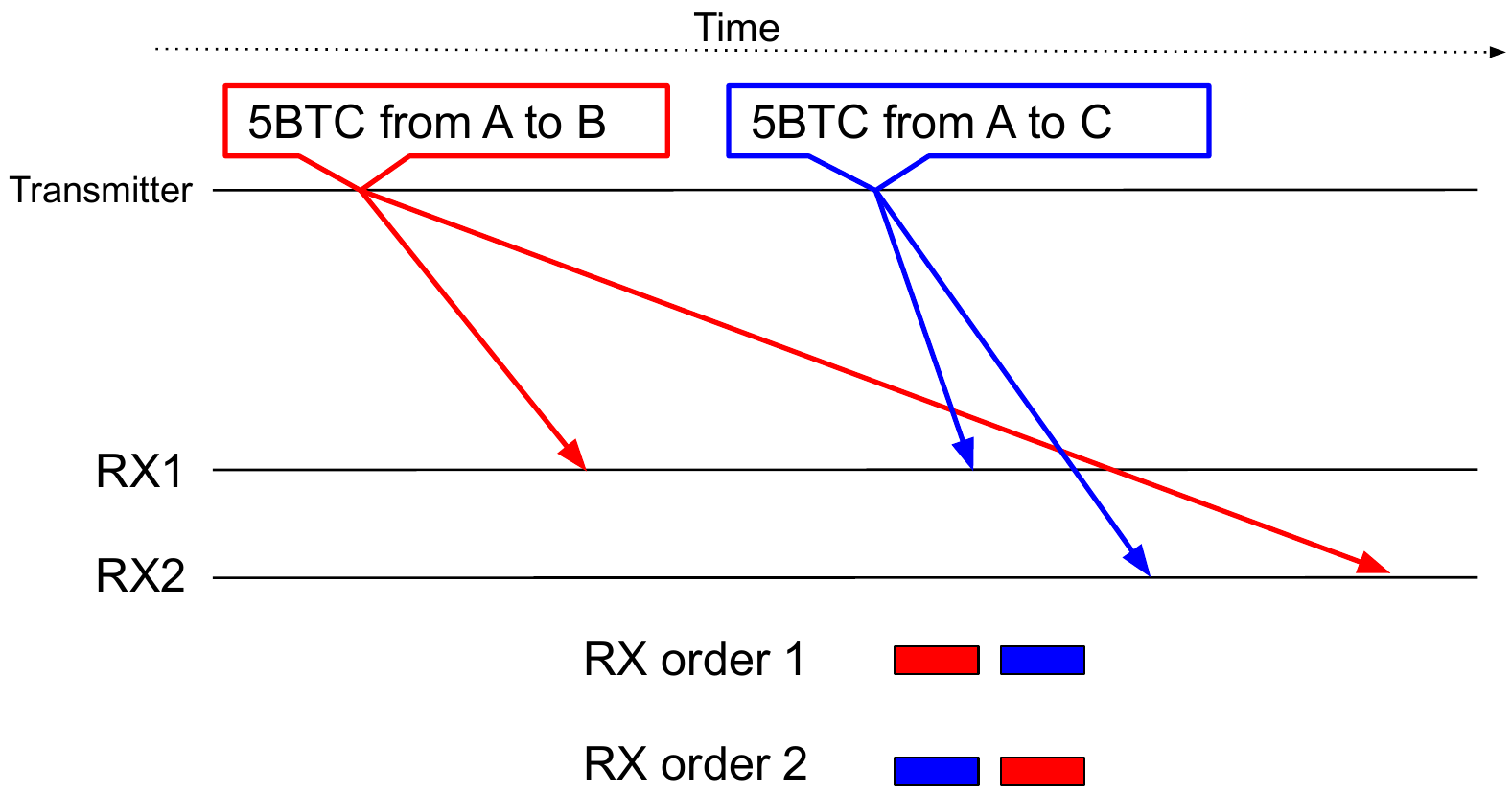} 
    \caption{Example of how propagation delays may lead to two different orders of reception at distinct validators.
Validators need to run a consensus protocol to find an agreement on the order of transactions. 
In this example if A owns only 5BTC then one of the two transactions must be rejected because it would represent a double spending.}
    \label{fig:dspendingTXorder}
\end{figure}

Two transactions that spend the same resources may be processed in different order by distinct validators spread across a P2P network.
This is because of different propagation delays in the network (\cref{fig:dspendingTXorder}).  
At this point, it is crucial for validators to find an agreement on the order of transactions
to determine which of the two came in first, and should be considered valid, and which came in second and should be rejected.

This fundamental problem is also known as \textit{Distributed Consensus Problem}. 
An equivalent problem is the implementation of a distributed timestamp server able
to sort transactions in an indisputable chronological order. 
The blockchain has been introduced in 2008 by the mysterious author of Bitcoin, Satoshi Nakamoto,
exactly as a way to implement a distributed timestamp server that assigns timestamps to
(blocks of) transactions, this way establishing their history.

However, the history of transactions may not be enough for a correct validation.
Indeed, a malicious user can alter the content of a block to repudiate an unwanted transaction,
ultimately falsifying the validation procedure.
To fend off falsification attacks a blockchain must be:
\begin{itemize}[itemsep=0em, leftmargin=0.5cm]
  \item \textit{Tamper-proof}: i.e., made so that is easy to verify that the registered transactions have not been manipulated after
  their recording, and it should be likewise easy to determine if these have been actually altered in a second instance of time.
  \item \textit{Immutable}: a blockchain-based ledger should adequately word off tampering attacks.
\end{itemize}
The tamper-proof property of blockchains is achieved by a clever embedding of \acp{CHF} into the blockchain data structure,
as explained in \cref{subsec:blockchain}.

\subsection{The Blockchain Data Structure}
\label{subsec:blockchain}

\begin{figure}[ht]
  \centering
    \includegraphics[width=\linewidth]{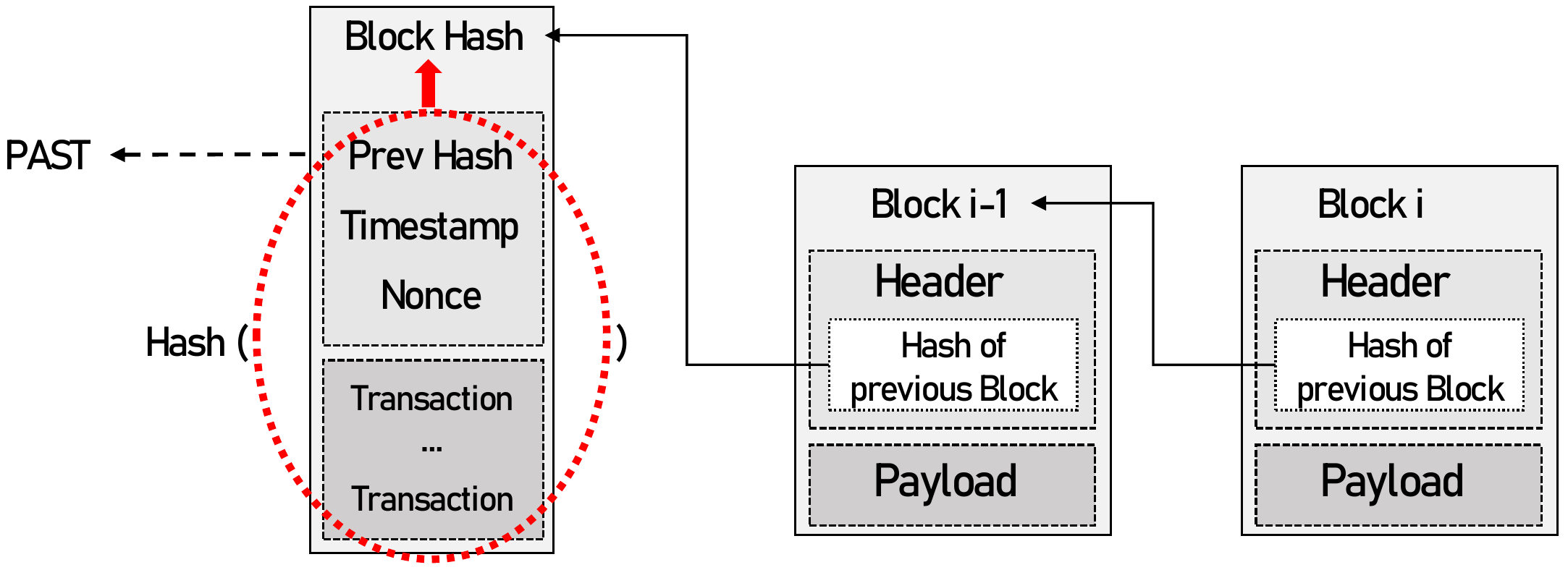} 
    \caption{The structure of a blockchain.}
    \label{fig:fblockchain}
\end{figure}

\acp{CHF} are crucial to make blockchains tamper-proof.
To become tamper-proof, transactions must be grouped into blocks including few other information as shown in \cref{fig:fblockchain}.

For example, in Bitcoin a block is valid if, hashing its content, the produced fixed-length digest%
\footnote{\ A message digest is a fixed size numeric representation of the contents of a message, as computed by a hash function.} %
exhibits a predefined number of leading zeros. This digest is said to be the ``Block Hash.''
The Block Hash must be a number lower than a given target, a target that can be changed to adjust
the difficulty~\cite{BitcoinDifficulty} of finding a valid Block Hash.
Finding a valid Block Hash can become an hard problem considering the random nature of \acp{CHF} and
the nodes strategy for generating new blocks, which is the following.

At first, a validator groups together some recent transactions and assigns them a timestamp.
Then, to further enforce the time dependency between blocks, it includes in the new block
the Block Hash of the last block he is aware of.
The references to previous blocks constitute the \textit{chaining} of blocks. 
Finally, a validator guesses a random value (the \emph{nonce}),
includes it in the new block, and applies the hash function to all these information
to compute a new digest, which becomes a candidate Block Hash.
This digest can be smaller than the target, thus not valid.
A node is usually forced to retry with as many different random nonces as possible,
until it produces (via brute force) a valid Block Hash. 

If someone tries to tamper block $i$, the attack will invalidate its Block Hash with high probability.
Because of blocks chaining also block $i+1$ gets invalidated and,
with a domino effect, all blocks following block $i$ get invalidated as well.
This mechanism makes the blockchain a \textit{tamper-proof} technology.

\subsection{\acf{PoW}}\label{subsec:PoW}
Consider that the \ac{CHF} mandated by the Bitcoin protocol is double-SHA256, which produces digests of 256 bits and,
at the current Bitcoin difficulty level, the first~77 bits must be zero.
The probability for a random nonce to be valid can be approximately computed as a function of the required number of leading zeros, that we call~$Z$. 
The set of all possible digests that the double-SHA256 function can generate has a cardinality of $2^{256}$.
Only digests that have $Z$ leading zeros are valid, so the cardinality of the set of valid digests is given by $2^{256-Z}$.
The probability $P(n)$ for a random nonce to produce a valid digest is therefore: 
\begin{equation}
  P(n) = \frac{2^{256-Z}}{2^{256}} =  \frac{1}{2^Z}.
\end{equation}
\noindent
For $Z=77$, then $P(n) \approx 6.62 \times 10^{-24}$.

When a node finds a valid nonce,  it can show it to all other nodes in the P2P network as a \textit{proof of work} (PoW), i.e., as a proof of the effort (computing power and ultimately energy in this case) that this node has spent to find such nonce. 
By showing a valid nonce, the node can claim the reward that the Bitcoin protocol assigns to nonce discoverers, which is the right of including a transaction that generates new Bitcoins and transfers them to the discoverer digital wallet. 
These rewards incentivize nodes in the hard task of discovering the very rare valid nonces. Those nodes that are constantly at work looking for valid nonces are metaphorically called ``miners.'' Asking miners to produce a \ac{PoW} for building a valid block is an important mechanism to
control the generation interval of new blocks (\cref{subsec:freqDifficulty}) and
and to secure the blockchain (\cref{PoWsecurity}).

\subsection{Block Generation Interval}\label{subsec:freqDifficulty}

\begin{figure}[htb]
  \centering
    \includegraphics[height=5.2cm, width=\linewidth]{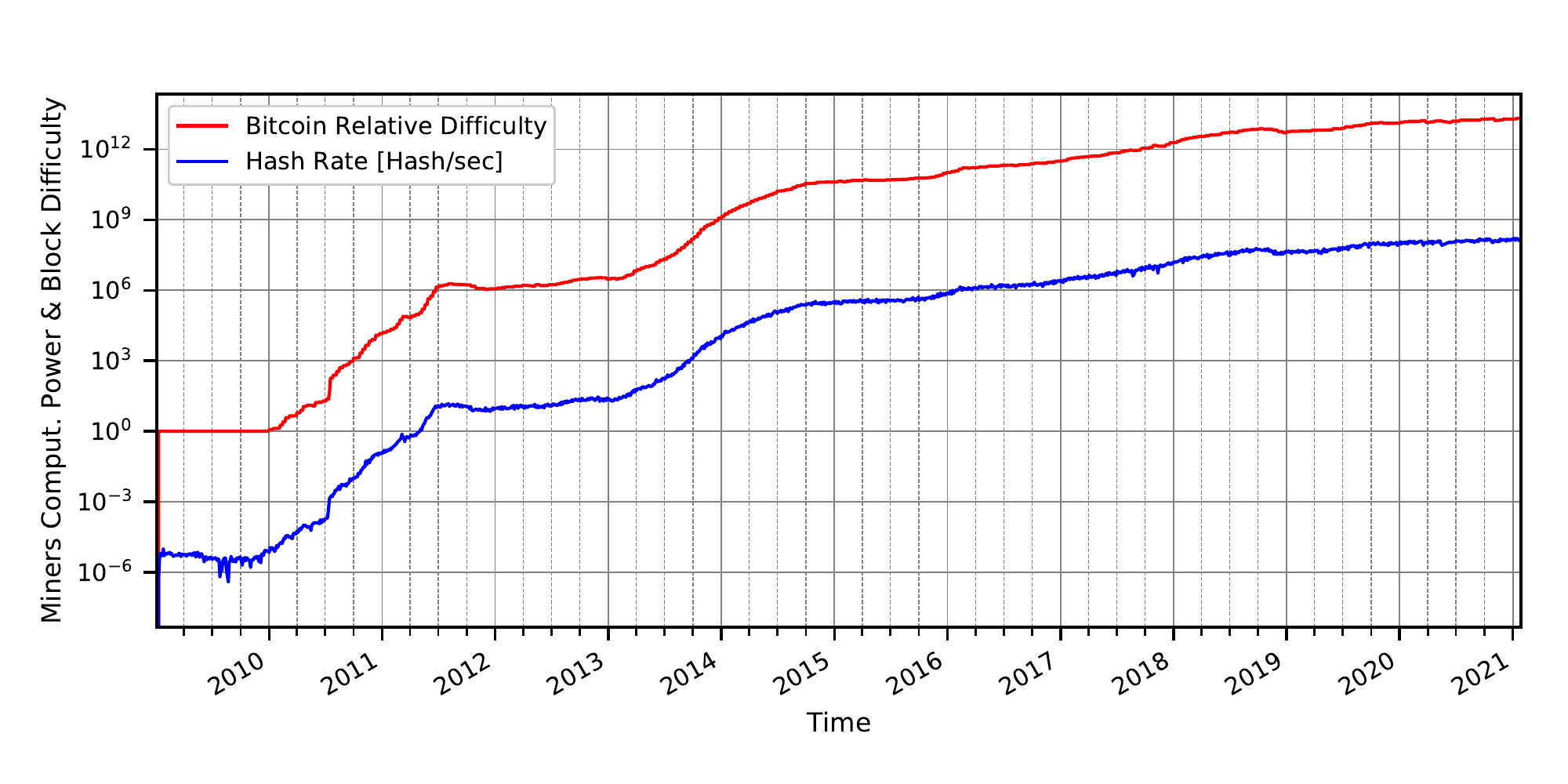} 
    \caption{Evolution of the Bitcoin network computing power, measured in hash per second (shown in logarithmic scale). 
    Over time the block difficulty has been adjusted to keep a constant average block production rate. 
    Statistics are taken from \protect\url{blockchain.com}}
    \label{fig:hashrate}
\end{figure}

The difficulty of the \ac{PoW} is tuned  so that, in the whole P2P network of miners,
a valid block is produced on average every 10 minutes. 
To keep a constant Block Generation Interval (\blockinterval), the difficulty of the \ac{PoW} must be tuned
according to the miners computing power, which has remarkably increased during the years (\cref{fig:hashrate}),
reaching the record of~166\,658 millions of Tera-hash computed per second (\terahash/s).

The \blockinterval must be kept high to avoid the simultaneous production of two blocks as far as possible.
Two blocks are considered ``simultaneous'' if the second one is generated within the 
average Block Propagation time (\blockprop), which for any P2P overlay based
on transactions is in the orders of seconds to maximum tens of seconds~\cite{Decker2013}.
Simultaneous blocks are problematic because their almost contemporaneous proposal divides the nodes
of the Bitcoin network in two parties that will append the two different blocks to the blockchain, forming two branches.
This situation can be represented by a bifurcation of the blockchain and is called a ``\textit{fork}.'' 
When a fork occurs, it means that there is no distributed consensus on block order anymore,
thus there is no agreement on the order of transactions. 
Without this agreement, the system is exposed again to Double Spending attacks. 

Usually forks are transient and are cleared as soon as another block is presented, making one branch longer than the other. 
This mechanism is called ``\textit{the longest-chain rule}''~\cite{courtois2014longest,shi2019analysis}, and it is inherent to the blockchain technology, thus representing a constraint in the design of blockchain based systems. 
The rule also implies that orphan blocks that do not end up to be part of the longest chain are not valid:
these blocks are also called \emph{stale blocks}.
The transactions included in the stale blocks, reward as well, will not be considered valid.
Miners do not want to waste resources working on blocks that will not be part of the longest chain, 
therefore, they immediately switch to a longer branch as soon as they notice such one,
increasing this way their chances of winning a more secure reward~\cite{courtois2014longest}.
This ensures that the majority of miners always work on the longest branch.

\subsection{The Security of the \ac{PoW}}\label{PoWsecurity}

While block hashing and chaining make blockchains tamper-proof, as explained in \cref{subsec:blockchain},
the \ac{PoW} is key to make blockchains \emph{immutable}.
Consider a malicious user that performed a transaction spending a considerable amount of cryptocurrency. 
This transaction was included in a given block, say, block~$i$, and the user now wants to revoke it.
One way to revoke this transaction is to cancel it from block~$i$, but this way
block $i$ would get invalidated and, by hash chaining, also all the following blocks: the tamper-proof
property of blockchains defuses this kind of attack.
Another strategy exploits the longest-chain rule and consists in generating a second
and longer chain of blocks that, starting from block $i-1$, would replace the current chain that contains block $i$.

\begin{figure}[htb]
  \centering
    \includegraphics[width=\linewidth]{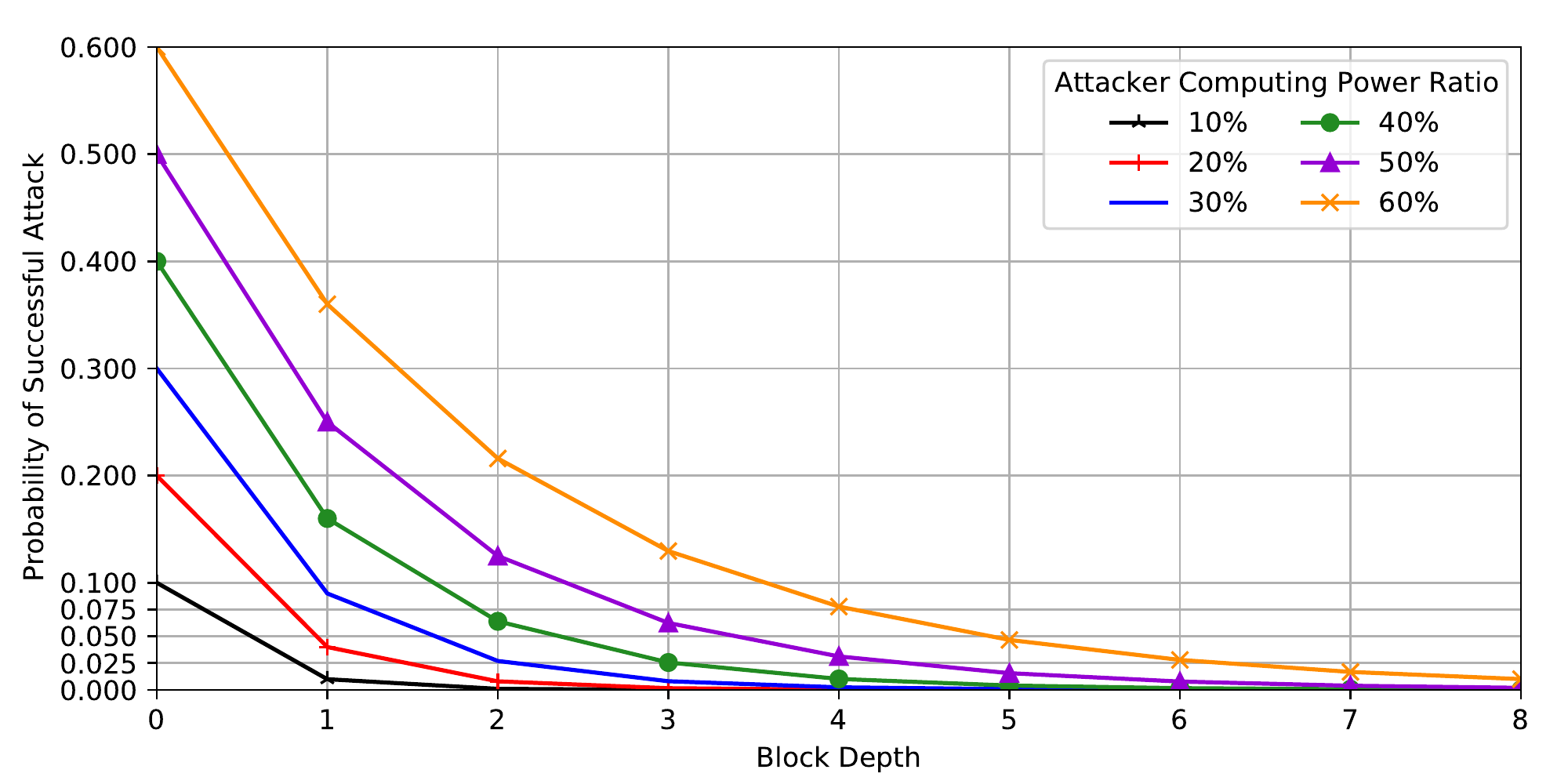} 
    \caption{Probability for an attacker that owns different amounts of
    computing resources to tamper a block, as function of the block depth.
    The depth of a block~$b$ is the number of blocks that have been added to the blockchain after $b$.
    The success probability decays exponentially fast, vanishing for blocks that are deep in the blockchain.}
    \label{fig:attackDifficulty}
\end{figure}

This attack that exploits the longest-chain rule is clearly a daunting task.
To be successful, the malicious user must beat all the other miners that during the attack keep using their collective computing resources to extend the blockchain with new blocks.
In general, the average probability $p_{a}$ of being the first to create and add a new block is the fraction of the computing resources controlled by a user, and block mining can be considered independent, so that the probability 
of adding~$N$ consecutive blocks is $p_{a}^N$, as shown in \cref{fig:attackDifficulty} for different computing power ratios.%%% 
%%%
\footnote{\ For each new block, all miners start competing to find a valid nonce almost at the same time, i.e., when the hash name of the last block is revealed and broadcast in the peer-to-peer network.
For this reason, all the ``races for the next block'' can be considered independent.}%
%%%

Some believe that owning the majority of the computing power ensures the control of the network through the so called ``50+1\% attack.''
The simple analysis above shows clearly that this is not enough, in general, 
to tamper any arbitrary block included in the history of transactions.
Rather, the existence of an extremely powerful user may lead to a biased
system where this user decides with high probability which transactions should be recorded
and which others should be ignored from now on in the future, but the immutability of the
deepest blocks of the blockchain is not harmed.
In fact, Bitcoin uses block depth as \textit{confirmation}, and recommends users to consider a transaction to be final
only if it has been confirmed by~6 subsequent blocks~\cite{confirmations}, a number that guarantees it is almost unassailable.
 
\subsection{Mining Overhead and Stale Rate}\label{subsec:staleRate}

The \emph{Average Mining Overhead ($\mathit{AMO}$)}
denotes the percentage of the network computing power that, on average,
does not contribute to the growth of the main chain, but rather leads to the generation of stale blocks. 
\begin{definition}[Average Mining Overhead]\label{def:miningOverhead}

\begin{equation*}
\miningOverhead := (1 - R) \times \frac{\blockprop}{\blockinterval}
\end{equation*}
\begin{itemize}[itemsep=0em, leftmargin=0.5cm]
  \item $R \in (0,1]$ is the ratio of the collective network computing power controlled
   on average by a successful miner. The complementary computing power ratio controlled by unsuccessful miners is $1 - R$;
  \item \blockprop is the average block propagation time;
  \item \blockinterval is the average block generation time interval.
\end{itemize}
\end{definition}
\noindent
\cref{def:miningOverhead} states that, in each block generation interval \blockinterval,
unsuccessful miners waste their computing power ($1 - R$) for \blockprop time, i.e.,
until they learn about the newly proposed block.
By definition, the AMO contributes to the generation of stale blocks.
Hence, it is proportional to the \emph{stale rate}, trivially defined by 
the ratio between the number of stale blocks and the total number of generated blocks.
\begin{definition}[Stale Rate]\label{def:staleRate}
\begin{equation*}
\frac{\#\text{Stale Blocks}}{\#\text{Stale + Final Blocks}}  \propto \miningOverhead
\end{equation*}
\end{definition}

The stale rate (\cref{def:staleRate}) is considered an indicator of the security level of a PoW-based blockchain,
because a higher stale rate means that a blockchain is more exposed to chain replacement 
and eclipse attacks~\cite{Gervais2016}.
Notice also that, fixing the other parameters ($R$ and \blockprop),
increasing the \blockinterval leads to a lower stale rate, enhancing the blockchain security.
The high \blockinterval (in the order of several minutes) chosen by many blockchains for cryptocurrencies~\cite{BitInfoCharts}
represents the effort of enhancing the mining efficiency and security by
trading transaction throughput and confirmation time.
Notice also that a higher \blockinterval enhances the miners profits, 
because reducing the mining overhead implies that investments on mining equipment become in percentage more profitable.

\subsection{Power consumption of the \ac{PoW}}\label{PoWpowerconsumption}

\cref{subsec:staleRate} highlighted how one can increase the security of a blockchain extending
the Block Generation Interval (\blockinterval).
An extension of the \blockinterval can be immediately achieved increasing the mining difficulty:
this way the cryptographical puzzle necessary to produce a valid block becomes harder, thus more time-consuming but also power-hungry.
The effort of setting up an exceptionally secure blockchain resulted, in Bitcoin, in a \ac{PoW}
that has become extraordinarily power-hungry~\cite{fairley2017ridiculous,deVries2018}.
The power consumption of the Bitcoin network in 2018 were estimated to be~2.55~GW, and forecast to reach~7.67~GW in the future.
This requirement is comparable to the energy demand of a whole country such as Ireland~\cite{deVries2018},
so it is hard to think that a blockchain secured by the PoW could be integrated in the
constrained domain of the IoT.

\section{Distributed Consensus Protocols}
\label{sec:distr-cons}

\cref{sec:background} posed the distributed consensus problem on the order of transactions and
explained how the PoW solves it.
The PoW advantages are many: it is extraordinarily secure, fully distributed, and user-agnostic.
In fact, users can participate to a PoW-based consensus without registering their identity
with some trusted registrar or bank, but just providing some computing power. 
Ultimately the PoW i) protects the user privacy and ii) free users from trusted authorities.
The popularity of blockchains, above all with cryptocurrencies, is most probably grounded in these two key aspects.
However, the PoW imposes also serious limitations in terms of transaction latency, throughput and power consumption.

It can be observed that consensus protocols are a crucial component for a Shared Ledger:
performance, consistency, policies of governance, security, and tolerance to failures are all properties of a Shared Ledger
that depend on the selected consensus protocol, rather than on the data structure used to record transactions.
A question arises: Is it possible to design a consensus protocol that preserves the PoW advantages and,
at the same time, avoids its drawbacks so to meet the typical requirements of IoT applications? 

The rest of this section revises the general limits of distributed systems, which constitute theoretical bounds for
the design of consensus protocols in general, and especially for IoT applications.
We first summarize these fundamental theorems valid for any distributed system (\cref{subsec:limitsConsensus}), 
then we explore the trade-offs inherent to the blockchain technology (\cref{subsec:trilemma}).
Finally, in \cref{subsec:shortRevConsensus}, we briefly review consensus protocols in search of alternatives to the \ac{PoW}.

\subsection{Limiting Theorems for Consensus}\label{subsec:limitsConsensus}

It is well known that it is impossible to achieve consensus in distributed systems in the presence of faulty nodes and unreliable communication channels \cite{FLPimpossibility}. 
The proof is based on this intuition: every time a consensus is close to be achieved among distributed agents, then a node or a communication failure may occur preventing the termination of consensus forever. 
This impossibility proof is in tight relation with another fundamental pillar of distributed systems, i.e.,
the \textit{Consistency, Availability} and \textit{Partition tolerance} (CAP) theorem\,\cite{CAPtheorem}. 
The CAP theorem states that whenever a system gets \textit{P}artitioned, then only two options are available: i) grant \textit{C}onsistency by safely blocking the system to fix the failures; or ii) keep processing transactions favoring \textit{A}vailability, with the risk that the two conflicting transactions (e.g., Double Spending ones) could be recorded, one per partition. 
This theorem is usually illustrated with a triangle with vertexes occupied by the three properties, stating that only two out of the three properties can be satisfied at the same time. 

Both the impossibility proof and the CAP theorem may be considered only mildly relevant, as they are valid only for ill-behaving systems, while in practice a system is built to work properly for most of its lifetime. However, they are the anteroom for the definition of two (almost equivalent) tradeoffs of tremendous practical importance. 

The first tradeoff is known as PACELC\,\cite{PACELC}, which advances the
CAP theorem (shuffling the acronym) and adding: \textit{Else Latency or Consistency}.
The novelty of PACELC is considering the case when the system is not partitioned, stressing on the tradeoff that arises between latency and consistency.
%%%
\footnote{\ Recall that the Bitcoin protocol enforces consistency introducing, by design, an average latency of 10 minutes per block, and further recommends to consider a block as ``unconfirmed'' until it becomes 6 blocks deep (\cref{subsec:blockchain}).}
%%%
A corollary to PACELC restricts the tradeoff to non-commutative transactions\,\cite{nonCommutativeTX}.
The observation that enables this corollary is that if two non-conflicting (commutative) transactions are performed in two different partitions, the overall consistency of the system is not compromised. 
This is an important restriction because, if transactions could be defined to be always commutative (which is impossible in a partitioned system), then an always consistent and available distributed system could be designed. 
An interesting consequence is that, if there are only few non-commutative transactions, all others can be executed in parallel to improve performance without sacrificing consistency.
The idea to define partitions of the systems that can process subsets of commutative (non-conflicting) transactions is also known as \textit{Sharding}, and empowers some of the most scalable permissionless blockchain solutions, e.g., \textit{Chainspace} and \textit{Omniledger}\,\cite{al2017chainspace,kokoris2018omniledger}.

\subsection{The Blockchain Trilemma}\label{subsec:trilemma}

\begin{figure}[htb]
  \centering
    \includegraphics[height=5cm]{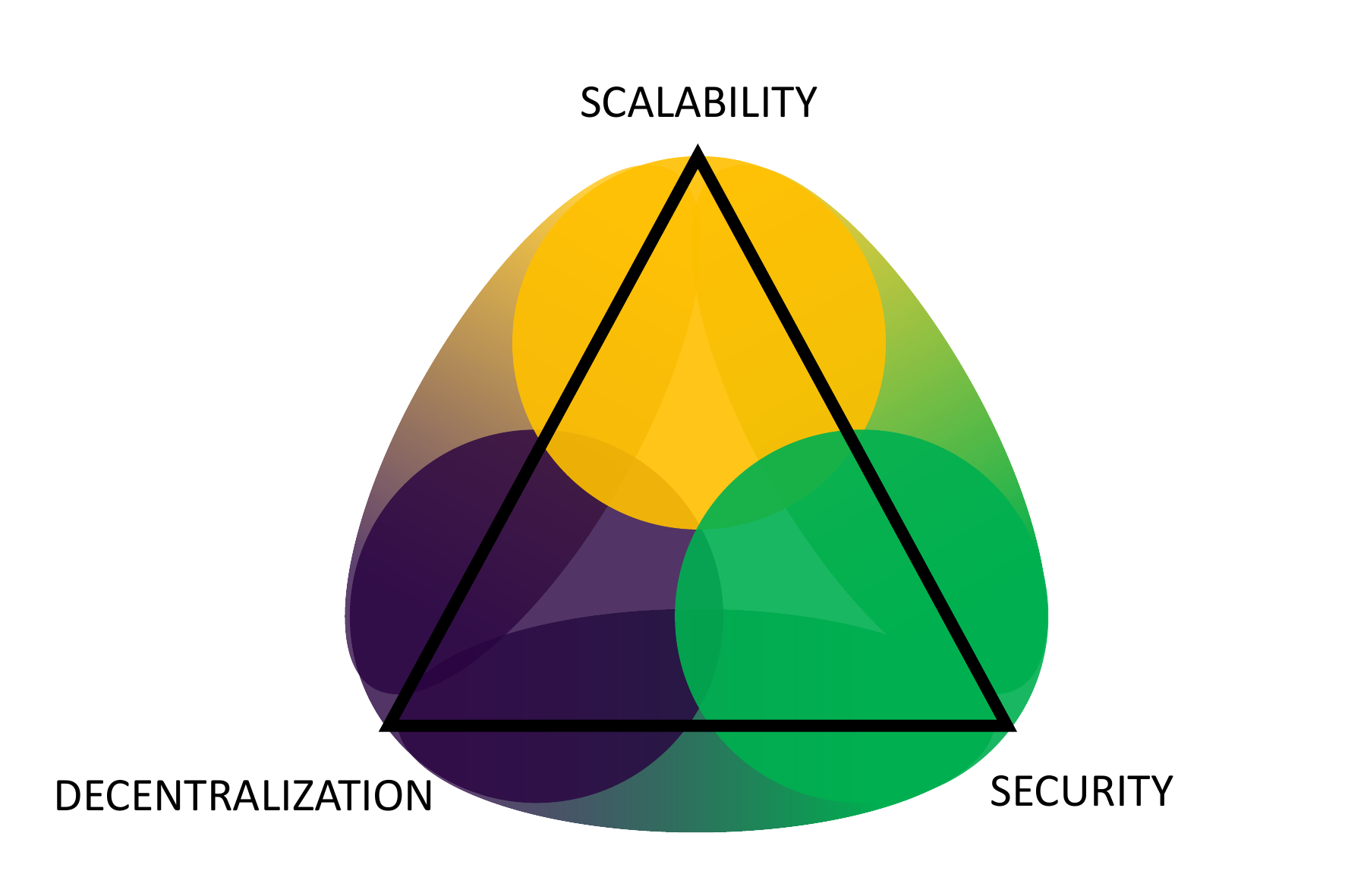} 
    \caption{The blockchain trilemma is illustrated by a triangle like the one in this figure
where one cannot draw a single point that is close to all the triangle corners, meaning
that a trade-off among the three properties must be chosen~\cite{grabe2020}.}
    \label{fig:trilemma}
\end{figure}

The second tradeoff is known as the \emph{``blockchain trilemma''}, illustrated in \cref{fig:trilemma},
which is essentially the reformulation of the PACELC theorem for the blockchain domain.
In particular, the trilemma illustrates the conjecture that a blockchain system cannot exhibit
maximum decentralization, security and scalability (performance) at the same time. 

Limited by the trilemma, an IoT developer willing to improve the network scalability may chose a consensus protocol less expensive than \ac{PoW} or reduce the mining difficulty to speed up the block generation rate. However, this would compromise the security, since less computing power becomes sufficient to perform a successful attack.
Another strategy could be to change the trust model, for example, restricting the access to the blockchain only to trusted, registered users.
This is fundamentally the strategy adopted with permissioned ledgers, in which a central registrar is introduced to authenticate users, but in this case decentralization is traded for a performance gain. 
Again, a trade-off must be chosen, as the trilemma warns us that no consensus protocol can 
ensure full security, decentralization and scalability~\cite{grabe2020}.

An IoT developer should therefore choose a consensus protocol and a blockchain-based system
only after having clearly identified the application requirements, choosing the most appropriate trade-off. 
A brief review of consensus protocols is reported in in \cref{subsec:shortRevConsensus}
to ease the selection of a consensus protocol for the IoT.

\subsection{Brief Review of Consensus Protocols}
\label{subsec:shortRevConsensus}

Consensus protocols are commonly partitioned into two broad families: Voting and lottery-based protocols. 
A third category is introduced in this short review to classify those protocols that elude this general dichotomy. 
\cref{tab:consensusComparison} is reported while concluding this section and compares the surveyed consensus
mechanisms in terms of scalability, security and decentralization. 

\subsubsection{Voting based Protocols}
Consensus protocols address the metaphor of the \textit{Byzantine Generals problem}~\cite{lamport1982byzantine}, i.e., the challenge for an ensemble of commanders to coordinate to perform a successful attack despite the potential betrayal of messengers, where betrayals model nodes/links failures and malicious attacks.
In classical solutions to this challenge, there is a node playing the role of ``leader'', which multicast a transaction to all other nodes, which in turn try to simulate the transaction locally and send back acknowledgments about the status of the operation. 
If the local transaction simulation completes with success, then the node replies with an acknowledgment suggesting to commit, otherwise to abort. 
Acknowledgments can be seen as votes.
As soon as the leader collects a trusted majority of votes for either commit or abort, then the decision is taken and broadcast back to all nodes.

\paragraph{\normalfont\textit{Leader-based and PBFT variants}}
The most popular protocol implementing this scheme is the \ac{PBFT} protocol \cite{PBFT}. 
It tolerates  up to $f$ malicious entities in a system with $3f + 1$ nodes, this is obtained requiring a quorum threshold of $2f + 1$ that ensures a majority of working/honest nodes. 
Many variants exist, they vary in terms of number of voting phases, mechanism to elect or change the leader and quorum thresholds. 
Well known examples of consensus protocols based on voting are the classic 2- and 3-PC \cite{bernstein1987concurrency} or PAXOS \cite{lamport2001paxos}, and many other more recent proposals 
\cite{ongaro2014search,aublin2013rbft,abraham2018hot,Androulaki2018,kwon2014tendermint, schwartz2014ripple}.

Leader-based and \ac{PBFT} variants are protocols that typically block (i.e., stop the decision process) when communication is asynchronous, where the keyword \textit{asynchronous} is used to represent all the reasons for a communication delay (the message arriving too late) or failure, such as unreliable channels, faulty or even malicious nodes. 
By blocking, safety is ensured at the cost of an increased latency, while progress (liveness) is granted only 
when the network recovers from the transient asynchronous phase. 

The potential blocks due to the failure of the leader are apparently inescapable. 
A unique leader must be part of the design; otherwise, a unique order of transactions cannot be defined.
This observation leads to the conjecture that the distributed consensus problem is equivalent to the problem of unique leader election \cite{gray2006consensus}.
In fact, leader election is harder than consensus.%
\footnote{\ \emph{``The weakest failure detector needed to solve Election is stronger than the weakest failure detector
needed to solve Consensus''}~\cite{leaderHarder}.} %
Unfortunately, the need of a leader comes with severe consequences~\cite{electionChapter,leaderConsiderations}. 
First of all, a single point of failure is introduced, so that it is sufficient to attack the leader via \ac{DDoS} to block consensus.
Secondly, the decisions taken by a single leader are not validated by any peer: 
having a leader becomes therefore a concern for the protocol fairness. 

\paragraph{\normalfont\textit{Pure Voting}}
In a pure voting system, each node sends its vote to all others, letting everybody perform the counting operations locally.
Running a voting phase in a pure voting system with $N$ nodes means that each of the $N$ nodes will send a vote to all other $N-1$ nodes, for an overall $\mathcal{O}(N^2)$ communication complexity. 
Pure voting systems avoid single point of failure, but their quadratic complexity prevents their deployment at the scale of an IoT network.

\paragraph{\normalfont\textit{Federated Voting and Sharding}}

Pure voting systems are impractical while leader based protocols, albeit efficient, introduce a single point of failure.
An hybrid approach can mitigate these issues. 
With Federated Voting or Sharding the network is partitioned and the consensus protocols are decomposed in smaller subproblems, solved within the federated enclaves (shards).
The Stellar protocol \cite{stellar} is a well-known representative of this class of protocols.
With Stellar, each node independently selects a set of trusted nodes (e.g., trusted  because of neighboring relationships) sets its own quorum threshold, then, applying a recursive principle of message passing and quorum overlapping, consensus can be achieved with a gossip protocol \cite{montresor1999gossip} in the whole federated network.
Gossip protocols normally require to work in cycles to guarantee convergence, introducing once more the problem of asynchronous networks. 

\paragraph{\normalfont\textit{Virtual Voting}}

Gossip algorithms are also the key feature of \textit{virtual voting} protocols, such as the hashgraph consensus algorithm~\cite{hashgraph}.
The idea is to let nodes spread transactions epidemically, attaching metadata about which received transactions 
have been critical to trigger a new transaction. 
At convergence, each node will own the full causal relationship between transactions, thus each node will be able to causally sort them, solving the consensus problem on their order.
A causal consistency criteria~\cite{Ahamad1995} is well known to be weaker than sequential consistency~\cite{lamport1979make},
but improves on latency by better supporting non-blocking recording of transactions~\cite{ojdb2015}.
The keyword \textit{virtual} is used because votes are not explicitly broadcast in an all-to-all fashion as described for pure voting systems,
still, all nodes implicitly acquire sufficient knowledge to sort transactions.

\begin{table*}[]
\centering
\caption{Comparison of Consensus Techniques}\label{tab:consensusComparison}
{\small 
\begin{tabular}{m{.13\linewidth}@{}m{.27\columnwidth}@{}m{.31\columnwidth}@{}m{.18\linewidth}@{}m{.44\columnwidth}@{}m{.32\columnwidth}}
\toprule
 &  \textit{Transaction Rate} & \textit{Network Scalability} & 
 \textit{Consensus Participants} & \;\textit{Attacks} & \textit{Centralization} \\ \midrule

\textbf{Leader-based} & High & High $\sim \mathcal{O}(N)$ & Registered nodes & \begin{tabular}[c]{@{}l@{}}Collusion of 1/3+1 nodes\end{tabular} & Central coordinator \\\midrule

\textbf{Pure Voting} & Low & \begin{tabular}[c]{@{}l@{}}Low $\sim \mathcal{O}(N^2)$\end{tabular} & Registered nodes & \begin{tabular}[c]{@{}l@{}}Collusion of 50\%+1 nodes\end{tabular} & Fully distributed \\\midrule

\textbf{Sharding} & Medium/high & Medium & Known neighbors & \begin{tabular}[c]{@{}l@{}}Colluded strategic minority\end{tabular} & Federations \\\midrule

\textbf{Virtual Voting} & Medium/high & Medium & Known neighbors & Neighbors can cheat & Fully distributed \\\midrule

\textbf{PoW} & Low & Low & Anonymous & \begin{tabular}[c]{@{}l@{}}Attack with control on\\huge computing power\end{tabular} & Fully distributed \\\midrule

\textbf{PoS} & Medium & Medium & Anonymous & \begin{tabular}[c]{@{}l@{}}Collusion of the richest\end{tabular} & Fully distributed \\\midrule

\textbf{DPoS} & High & \begin{tabular}[c]{@{}l@{}}High\end{tabular} & Elected Delegates & \begin{tabular}[c]{@{}l@{}}Collusion of delegates\end{tabular} & Requires delegates \\\midrule

\textbf{Round-Robin} & High & \begin{tabular}[c]{@{}l@{}}High\end{tabular} & Registered nodes & \begin{tabular}[c]{@{}l@{}}Block by any malicious user\end{tabular} & Requires global synch \\\midrule

\textbf{node-to-node} & High & High & Known neighbors & Neighbors can cheat & Fully distributed \\ \bottomrule
\end{tabular}
}
\end{table*}
%%%%%%%%
\subsubsection{\textbf{Lotteries and Proofs}}\label{subsubsec:lotteriesProofs}
So far the consensus problem has been solved by voting, with votes used to elect a leader that imposes his decision or to form
a majority in favor of a particular action.
Another way to address the consensus problem is to organize a lottery, and the lottery winner becomes the leader.
The winner must provide a sort of winning ticket, a ``proof'' to be shown to claim the consensus. 
All of the consensus protocols that resemble a lottery game are associated with a specific proof
that is required to lead the consensus. 

\paragraph{\normalfont\textit{Proof of Work (PoW)}}
For instance, in \cref{subsec:blockchain} the \ac{PoW} has been introduced and the nonce is the winning ticket that miners need to show to propose a block, imposing their order of transactions.
Albeit effective to achieve consensus, the drawbacks of \ac{PoW} are evident. 
First of all, it slows transactions speed, but also results in a considerable waste of computing power and lack of fairness, as long as miners are free to include the transactions they prefer in the block they are trying to forge, thus can deliberately ignore the transactions of competitors. 

\paragraph{\normalfont\textit{\acf{PoS}}}
The \ac{PoS} is an energy-aware alternative to \ac{PoW} that relies on economical rationality to achieve consensus.
In \ac{PoS}, a randomized process selects a leader, and the key property of the random process is to bias those entities that own more cryptocoins, or whatever resource is at stake.
The reasoning is that the owners of many cryptocoins (the richest stakeholders)
have a vested interest in keeping the network working well and trusted, so that the
system could be perceived as valuable, and the value of the owned cryptocoins is safeguarded and enhanced. 

\paragraph{\normalfont\textit{\acf{DPoS}}}
A variant of \ac{PoS} that recently became very popular is \ac{DPoS}, which is implemented for example by EOS, Tron, Steem, and Bitshares, and outperforms all other consensus protocols in terms of scalability \cite{bach2018comparative}. 
With \ac{DPoS} stakeholders vote to elect delegates, and their votes are weighted according to the fraction of owned coins. 
Sometimes delegates need to show commitment with a deposit (escrow) that can be confiscated if they do not run the internal consensus protocol honestly. 
The result is that delegates are chosen according to an economic rational criterion, and given that delegates are few and trustable (they are committed  and accountable), they can achieve consensus much faster. 

\paragraph{\normalfont\textit{Other Proofs}}
Many other lottery based protocols have been proposed in the last years.
They all evolve around the concept of proving commitment either by showing to be willing to sacrifice resources,
or by the fact that the user owns a considerable stake. 
For the sake of comprehensiveness, other known proofs are: 

\begin{itemize}[itemsep=0em, leftmargin=0.3cm]
  \item Proof of Elapsed Time (PoET) \cite{chen2017security}: where the sacrificed resource is the (random) time spent in a waiting queue;
  \item Proof of Importance (PoI)\footnote{\ This concept was introduced in NEM P2P cryptocurrency \url{https://docs.nem.io/ja/gen-info/what-is-poi}, but never published in a scientific location to the best of our knowledge.} and Proof of Networking (PoN) \cite{ghiro2018proof}: where the node commitment in the network is computed on the base of a different mix of metrics, including network topological information;
  \item Proof of Burn (PoB): a node must burn coins, sending them to a dead address, to gain the privilege of leading the consensus;
  \item Proof of Capacity (PoC): if memory is the main resource necessary to solve a cryptographical problem, then \ac{PoW} becomes PoC, like in \cite{larsson2018cryptocurrency};
  \item Proof of Deposit (PoD): with \ac{PoS}, nodes with ``nothing at stake'' can behave maliciously without any punishment. In Casper \cite{buterin2017casper}, entities have to deposit some coins that are confiscated in case of malicious activities.
\end{itemize}

\subsubsection{\textbf{Other Approaches}}

\paragraph{\normalfont\textit{Round-Robin}}
A straight-forward mechanism to address the distributed consensus protocol is to let all nodes succeed each other, in consecutive turns, as leader.
Full trust among nodes is required to fairly run the successions procedure, which can be blocked indefinitely by any malicious participant.

\paragraph{\normalfont\textit{Node-to-node Consensus}}
The \textit{node-to-node} keyword is used to indicate two ore more mutually
trusting neighbors that agree to privately perform transactions.
Usually these neighbors open a fast-payment channel to handle their frequent private transactions,
so that they avoid paying the multiple fees that would incur if a public blockchain were used instead.
Node-to-node transactions are not visible on a main blockchain, so they are said to be \textit{off-chain}~\cite{poon2016bitcoin,althea}.
Infrequently, the two nodes close their channel issuing a closing transaction reported on a main blockchain.
This closing transaction may represent the balance of thousands or more off-chain transactions: 
the off-chain transactions rate can hugely enhance the overall network transaction rate.

\section{Towards a Blockchain Definition}
\label{sec:demystify-blockchain}

We have discussed the structure of the blockchain (\cref{sec:background})
and distributed consensus protocols (\cref{sec:distr-cons}),
stressing on the limits and trade-offs inherent to the blockchain technology.
However, browsing the literature related to blockchain applications makes the blockchain seem as an almost universal, limitless technology.
The application range for the blockchain, in fact, seems to be so broad to include:
Supply Chain Management~\cite{montecchi2019s,tse2017,tian2016agri,miller2018},
E-Voting~\cite{bcEvoting,ayed2017,bistarelli2017,liu2017voting,hardwick2018voting,wang2018large,qi2017,noizat2015},
Smart-Grid~\cite{Gai2019,lombardi2018,mylrea2017,Mengelkamp2017,munsing2017,hahn2017smart,laszka2017,yan2017},
Healthcare~\cite{Xu2019,dwivedi2019,bcHealthcare,Yue2016}, Banking~\cite{Cocco2017,Guo2016}, Smart Cities \cite{Shen2019,ibba2017city,patil2017}
and even Vehicular~\cite{Liu2019,bcVehicular,Yang2019,patel2019vehichain,cebe2018,michelin2018,zhang2018vanet,li2018coin4Vehic,rowan2017,sharma2017,singh2017,yang2017,yuan2016,leiding2016} and Drone~\cite{kapitonov2017,liang2017,bcDrones,ferrer2016,bcUAV2} Networks,
going way beyond the original cryptocurrencies such as Bitcoin and Ethereum.
We argue that this apparent universality of the blockchain is rooted in the ambiguity of the \textit{blockchain} term itself.
This ambiguity complicates clarification, since the blockchain can have different roles for the IoT depending on the different possible ---if not improper--- definitions. 

To unriddle the possible usage of the blockchain we first need to formulate a connotative
and precising definition for the term ``blockchain.''
To this end, we compare the most popular platforms commonly considered as emblematic blockchains with standard \acp{DB},
looking for the distinguishing features that will constitute our blockchain definition.
Thanks to this definition we can elucidate the conditions for a proper use of a blockchain,
and we also provide guidelines for deciding when to avoid it and instead prefer to rely on traditional solutions. 

\subsection{Blockchain vs.\ Traditional Technologies}
\label{subsec:blockChainVStradition}

\begin{figure}[htb]
  \centering
    \includegraphics[width=\linewidth]{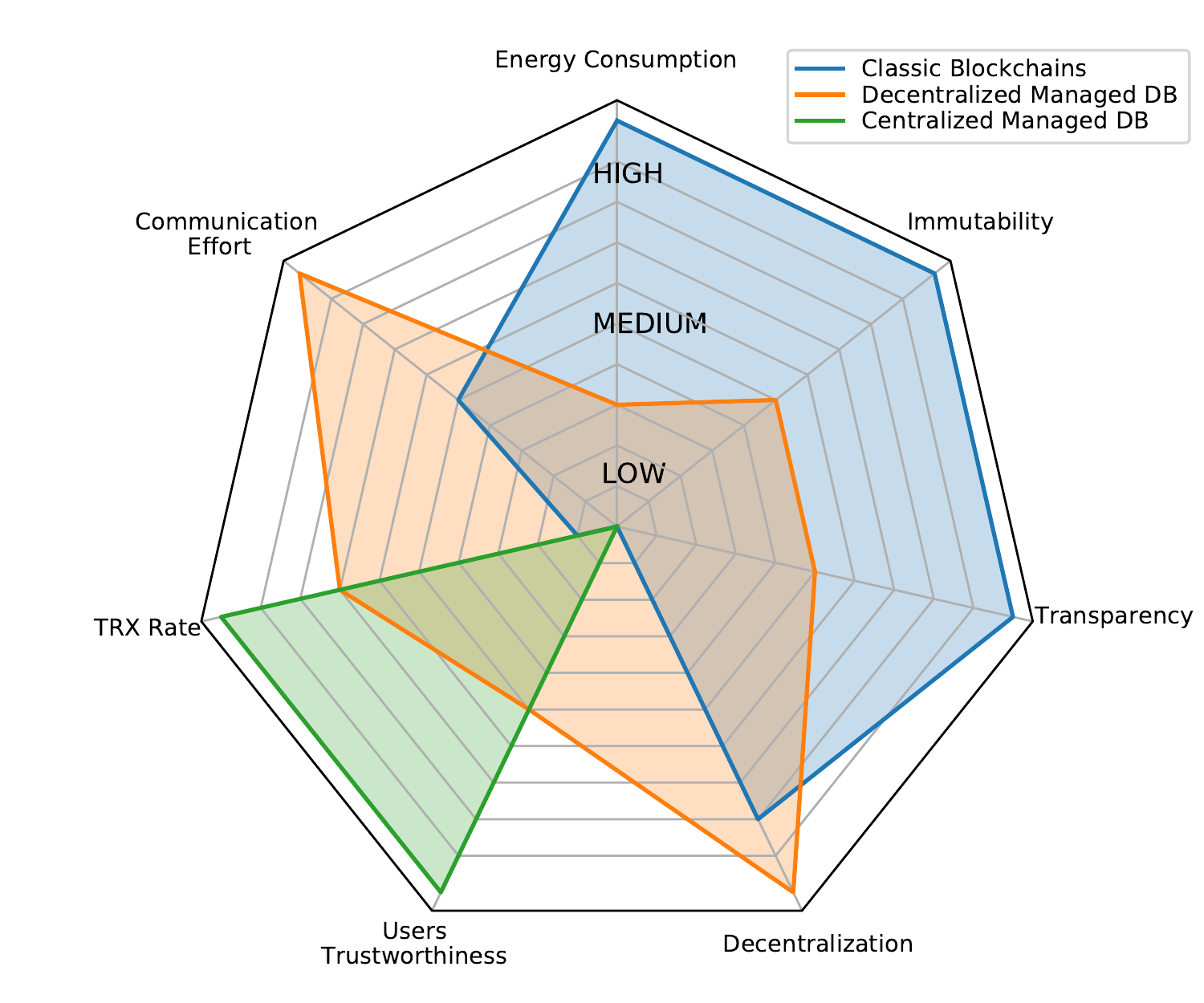} 
    \caption{Multidimensional comparison of the blockchain with traditional Shared Ledger technologies: i.e., centralized and decentralized managed \acp{DB}.}
    \label{fig:spiderchart}
\end{figure}

\cref{fig:spiderchart} aids the discussion about the technological advantages and disadvantages of
the competing technologies for the implementation of a Shared Ledger.
The first two are traditional \ac{DB} technologies, that will be compared with the still undefined concept of ``Classic Blockchains.''
Our intention is to capture under this concept all those platforms (such as Ethereum and Monero~\cite{monero})
that preserved the distinctive features introduced in history for the first time by Bitcoin,
marking a new era for the Shared Ledger technologies.

\paragraph{Centrally Managed \acp{DB}}
A centrally managed \ac{DB} is maintained by a central administrator, such as a trusted employee or
a single company in charge of keeping the \ac{DB} well maintained.
The recorded data can be shared among various clients upon request.
The central manager can, at his own discretion, authorize or deny the access to the \ac{DB}.
According to the described paradigm a centrally managed \ac{DB} represents a possible implementation of a Shared Ledger.

The greatest advantage of one such implementation is the high level of efficiency in terms of transaction rate,
communication effort and power consumption.
In fact, an administrator can exploit the decades refined technology of commercial \acp{DB} 
to achieve maximum performance at a low power consumption.
The administrator works autonomously, so it also avoids the communication efforts of a consensus protocol
that would become necessary to coordinate more \ac{DB} maintainers.
If advantages are many, disadvantages are numerous too. 
For example, the trust in the administrator must be absolute because
this administrator can in principle tamper, censor or even resell users data. 
A centrally managed \ac{DB} is not considered transparent as well, because
nobody controls nor validates the admin operations.
Similarly it cannot even be considered immutable, as the admin is free to delete data.

\paragraph{ Distributedly Managed \acp{DB}}
Distributedly managed \acp{DB}, i.e., \acp{DB} cooperatively maintained by a group of administrators,
represented the only option to implement a decentralized Shared Ledger before the rise of blockchains.
Redundant DB copies are introduced: nodes chose and run a consensus protocol to agree on writing operations,
enforcing this way a consistency model~\cite{lamport1979make,adve1996shared}.
This distributed architecture provides a varying degree of tolerance to failures
which depends on the strength of the consensus protocol and on the number of redundant DB copies. 
The price paid by distributed \acp{DB} for decentralization is the increased coordination effort
necessary to run the consensus protocol, that also slows down the transaction rate.
A distributed \ac{DB} is harder to tamper compared to a centralized one, since an attacker must corrupt more nodes. 
All write operations are validated by a quorum of peers: this mechanism enhances transparency
as no absolute trust in the admin is required anymore.
Nonetheless, the system is secure only if a majority of peers is honest. 
The maintainers of the distributed \ac{DB} are free to record data in any data structure (not necessarily a block-chain),
provided that the application requirements are enforced by other factors, e.g., through rules of the internal consensus protocol.

\paragraph{ Classic Blockchains}
Iconic blockchain platforms are Bitcoin (described in \cref{sec:background}) and Ethereum.
Both are based on the \ac{PoW}, precisely as many other popular \ac{PoW}-based blockchains%
%%%
\footnote{\ Examples of other famous \ac{PoW}-based cryptocurrencies are
Bitcoin-Cash, Litecoin, Namecoin, Dogecoin, Primecoin, Auroracoin, Monero, Etherum-Classic and Zcash.
For a more complete list of cryptocurrencies the reader can refer to~\cite{listCryptoCC}.} %
%%%
that, together, account for more than 90\% of the total market capitalization of existing digital cryptocurrencies~\cite{armknecht2017,coinmarketcap}. 
Some authors refer to these public, permissionless, \ac{PoW}-based blockchains as to \emph{classic blockchains}~\cite{vademecum}.

Classic blockchains turns out to be a particular case of decentralized \ac{DB} where transparency and immutability are constitutional and brought to their extremes.
The only data structure used in a classic blockchain is, unquestionably, a block-chain, i.e., a special linked list characterized by cryptographic links, and blocks of transactions as items of the list.
In a blockchain, data can only be appended and it is never deleted or modified. 
All append operations are public and transparent, so that the validity of all transactions can be verified at anytime by any peer. 
A classic blockchain is open to any anonymous user, therefore a very strong consensus protocol is necessary to safeguard  the ledger.
This leads to the well known drawbacks: slow transactions rate, high latency, and huge power consumption. 
Albeit slow, strong consensus mechanisms adopted in public blockchains allow the removal of trusted authorities and enable transactions also among anonymous, untrustworthy users.

\subsection{Connotative Definition of Blockchain}
\label{subsec:definition}

The analysis of the competing technologies for the implementation of a Shared Ledger (\cref{subsec:blockChainVStradition})
suggests what are the distinctive characteristics of a blockchain that distinguish it from all other \acp{DLT}.
We condense these characteristics in the following definition of the term ``blockchain'':

\vspace*{0.4em}
\begin{mydefinition}[label={BCdef}]{Characteristics of a Classic Blockchain}
 \begin{enumerate}[]
  \scshape
  \setlength\itemsep{0.25em}
  \item Openness to Anonymous Users
  \item Full \& Public History of Transactions
  \item Strong Distributed Consensus Protocol
\end{enumerate}
\end{mydefinition}

The {\scshape Openness to Anonymous Users} is the first, essential feature of a blockchain.
The blockchain ability to preserve the privacy of users more than what is done by banks or by
other centralized implementations of a Shared Ledger comes ultimately from the anonymity of users.
The openness to anonymous users is also fundamental for making blockchains decentralized.
If users had to be identified, a centralized trusted registrar ---potentially discriminatory---
would become necessary, compromising the ledger decentralization.
The openness to anonymous users is thus constitutional for a blockchain,
but introduces also a new problem about the disputation of transactions,
because it is not possible to prosecute an anonymous, untraceable user in case of fraud:
users must accept that transactions are, de facto, indisputable.

In the trustless scenario made of anonymous users, one can accept an indisputable transaction
only if it is empowered to perform, on its own, a complete check of validity of any transaction at any time.
Involving a trusted authority, such as a bank, this user could trust
the private and opaque internal ledger of this bank to manage transactions, but removing trusted intermediaries implies
the necessity of keeping a complete record of all transactions on a ledger open to the public,
otherwise the distributed validation of transactions becomes impossible.
As seen in \cref{needTRXhistory}, the solution offered by blockchains is to record 
the {\scshape Public \& Full History of Transactions}, so that anyone can verify 
that no previous transactions in the whole history already spent the resources being transacted.

Despite the availability of a public and complete ledger, the validation of a transaction can be still falsified
by an attacker that manipulated the history of this transaction, tampering the blockchain for gaining a personal advantage.
Therefore the ledger (i.e., the blockchain) must be safeguarded by a 
{\scshape Strong Distributed Consensus Protocol}, otherwise nobody would trust the system.
This is why a mechanism like the \ac{PoW} that, as explained in \cref{PoWsecurity},
makes historical blocks immutable and is mathematically secure regardless of any trust assumption on users,
is a distinctive element of blockchains.
This kind of immutability is so fundamental for the concept of blockchain that [sic] 
\emph{for cryptocurrency activists and blockchain proponents even  simply  questioning
the  immutable  nature  of  blockchain is  tantamount to heresy}~\cite{politou2019blockchain}.

To recap: a blockchain is designed specifically to guarantee full memory of all transactions open to any anonymous user to enable the distributed validation of transactions avoiding trusted intermediaries.
But the blockchain alone is meaningless without a mechanism that safeguard the 
historical records of transactions from tampering attacks, 
this is why a {\scshape Strong Distributed Consensus Protocol} becomes essential.
The arguments we used to justify our definition of blockchain are concisely summarized in \cref{prop:logicProof}.

\begin{highlightBox}[label={prop:logicProof}]{Arguments supporting \cref{BCdef}}
 \vspace*{-0.4em}
 Users Anonymity $\Rightarrow$ non-disputable Transactions;\vspace*{0.35em}
 
 If the Ledger is \_ then New Transactions are \_ :\vspace*{-0.3em}
\begin{itemize}
\setlength\itemsep{0.01em}
  \item Private $\lor$ Partial $\Rightarrow$ \textit{unverifiable}
  \item Public $\land$ Full $\Rightarrow$ \textit{verifiable}
\end{itemize}
\vspace*{-0.5em}
A Strong Consensus makes the Ledger immutable, protecting it from falsification. 
\end{highlightBox}

\subsubsection{Comparison with other definitions}
A first definition of blockchain, in computer science, can be restricted to the simple data-structure made of blocks of information
chained by hash-pointers, known in the literature since the '70s~\cite{hashLink,Halatsis1978}. 
However, we believe that the introduction of Bitcoin and Ethereum enlarged the meaning of the term blockchain.
As a matter of fact, Iansiti and Lakhani propose a wider definition which is the following:
\emph{``[The] blockchain is an open, distributed ledger that can record transactions between two parties efficiently and in a verifiable and permanent way''}~\cite{iansiti}.

This definition highlights the blockchain operational purpose as distributed ledger,
distinguished from other traditional ledgers because of its Openness, Verifiability and Immutability (permanent records) properties.
Our characterization highlights these same features but it does so stressing on 
the constitutional elements that connote a blockchain as an open, verifiable and immutable ledger technology.
\cref{BCdef}, in fact, characterizes the blockchain as a technology 
\textit{open} to any anonymous user,
\textit{verifiable} thanks to the complete and public recording of all transactions, and as much 
\textit{immutable} as possible by reason of a strong distributed consensus protocol. 
Our definition, compared to previous ones, is not axiomatic: rather, 
acknowledges the blockchain as an open, verifiable and immutable technology
deriving these properties from the three, inseparable elements that together constitute the essence of a blockchain.

Our definition results to be more restrictive, as it reserves the epithet ``blockchain'' only to those platforms
that fully reflect the blockchain specificity described by \cref{BCdef}.
The various platforms usually mentioned as blockchain without actually complying with \cref{BCdef} are criticized in \cref{critiquePermissioned}.

\cref{BCdef} also serves our clarification purpose,
being the base for claiming that \textit{the blockchain is not a universal technology},
but it rather has precise characteristics advantageous only for a limited number of applications.
The many applications proposed in the recent literature that exhibit features in contrast with
\cref{BCdef} are criticized in \cref{sec:pitfalls}.

\subsection{Permissioned Ledgers are Blockchains?}\label{critiquePermissioned}
\cref{BCdef} raises a question: since permissioned ledgers are not openly verifiable,
nor safeguarded by a strong consensus protocol, shall we call them blockchains?

\paragraph{ Not an Open, Decentralized, Verifiable Technology}
By definition, in permissioned ledgers the access is restricted only to permissioned users.
A central, trusted registrar responsible for the identification of users and for granting permissions
must therefore exist.
Moreover, permissioned ledgers are typically used by enterprises to record business-critical transactions,
that consequently are kept confidential and cannot be verified by an external agent.
For these reasons, permissioned ledgers are not a truly decentralized nor an open technology.

\paragraph{ Less Immutable means less Secure}
A trust model with registered and permissioned users is certainly safer
than one where the ledger is open to any anonymous user. 
Strengthen by stronger trust assumptions, permissioned ledgers usually abandon the secure but power-hungry \ac{PoW} 
replacing it with more traditional, efficient consensus protocols.
However, this way they return to be vulnerable to traditional attacks 
led by the ``simple'' --- i.e., ``inexpensive'', not discouraged by any costly sacrifice--- collusion of a majority of users.
Permissioned ledgers are therefore less immutable and less secure than iconic blockchains such as Bitcoin or Ethereum,
that instead are not affected by this vulnerability.

\paragraph{ Permissioned Platforms are Traditional Ledgers}
Permissioned platforms seem to be not much different from traditional ledgers that existed also before
Bitcoin~\cite{permissionedNOTbc}, as they are empowered by traditional consensus protocols and their trust model
still depends on a central authority, while permissionless blockchains such as Bitcoin revolutionized the state-of-the-art.
From an historical perspective permissioned platforms represents therefore only
a technological sophistication of the older, traditional \acp{DLT}.

\begin{figure}[h]
  \centering
    \includegraphics[width=\linewidth]{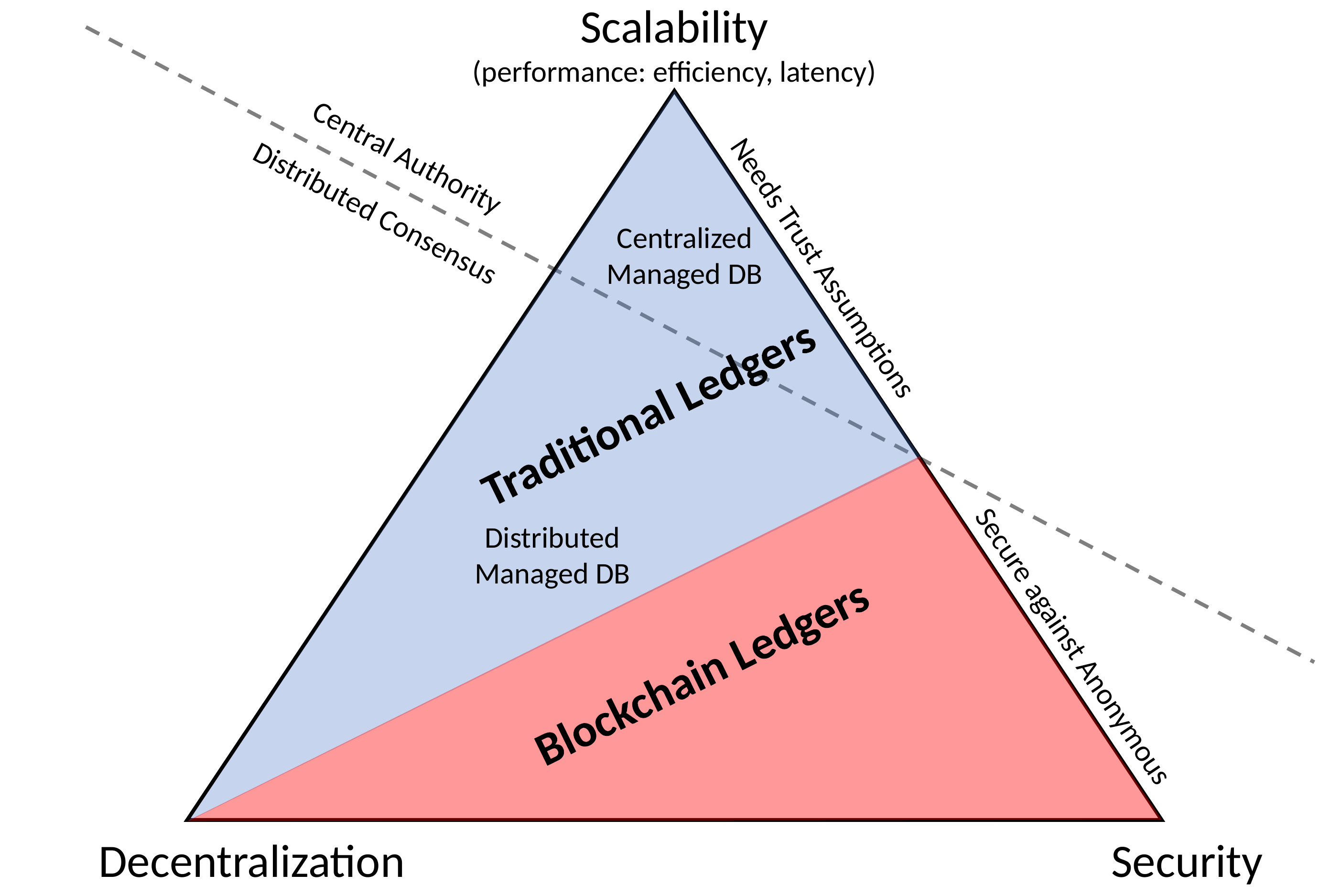} 
    \caption{Position of the Blockchain in the landscape of the Shared Ledger technologies.}
    \label{fig:landscapeSLtech}
\end{figure}

For these reasons, from now on we will consider permissioned platforms as belonging to the broader class
of traditional DB-based ledgers, rather than blockchain representatives.
\cref{fig:landscapeSLtech} depicts our vision of the landscape of Shared Ledger technologies, 
with the blockchain positioned according to \cref{BCdef} as provided in \cref{subsec:definition}.

\subsection{Proof of Work or Proof of Stake?}\label{PoWorPoS}
The transition from PoW to PoS planned by many popular blockchain systems (in primis Ethereum)
to stop wasting computing power can compromise the blockchain characteristics?
Many would be the reasons to dismiss the PoW:
\begin{itemize}[itemsep=0em, leftmargin=0.5cm]
  \item Enormous power consumption (\cref{PoWpowerconsumption} and \cite{fairley2017ridiculous,deVries2018});
  \item Modest transaction rate and high latency~\cite{Gobel2017,croman2016},
  if compared to traditional Byzantine-Fault-Tolerant alternative protocols~\cite{Vukolic2016};
  \item Tendency for the computing power to consolidate under the control of few large mining pools%
\footnote{\ The popular blockchain.com website reports the hashrate distribution
of the largest Bitcoin mining pools~\cite{BitcoinPools},
with the 12 largest of them controlling almost the 80\% of the network computing power.
Notice, however, that pools are consortia that aggregate together multiple miners,
so they still achieve a certain degree of decentralization.}~\cite{gencer2018,Gervais2014}. %
Moreover, under \ac{PoW} it may be advantageous for few powerful users to collude behaving as \emph{``selfish miners,''} again damaging the decentralization level of the network~\cite{eyal2014,Carlsten2016}.
\end{itemize}

For these reasons the transition to the PoS may be justified, however, also the PoS has been criticized:
\begin{itemize}[itemsep=0em, leftmargin=0.5cm]
  \item The PoS design deliberately priorities the richest stakeholders,
   these last tend to accumulate voting power damaging the network decentralization.
   This tendency is usually condensed in the motto \emph{``the richer get richer''}~\cite{fantiCompounding}.
  \item While there exist rigorous studies on the convergence and on the byzantine-tolerance
  of both traditional voting protocols and of \ac{PoW}-based blockchains~\cite{Gervais2016},
  the economics of stake-based systems suggest that their equilibrium is
  not always granted~\cite{catalini2019market,Harvey}.
  This raises the concern that, like with real markets, stake-based protocols will lead
  to unstable systems subject to market failures and bubbles~\cite{fanti2019economics}.
  \item A \ac{PoS}-based blockchain is reversible by means of long-range or stake-bleeding attacks~\cite{Deirmentzoglou2019,Gazi2018},
  thus its immutaiblity is considered questionable.
\end{itemize}

In light of the discussions on the limits of PoW and PoS we claim that,
essentially, they both empower a \textit{census suffrage} system. 
In the case of PoW only rich users that can afford the sophisticated mining equipment can participate in the protocol. 
In PoS a similar restriction on voting by census is directly embedded in the protocol,
with the remarkable advantage of saving a huge amount of energy, 
but with the risk of long term instabilities.
There is, however, a key difference:
while the acquisition of computing power is subject to natural factors such as 
the cost of electricity, the value fluctuations of a \ac{PoS}-system only depends on speculative mechanisms. 
So while with PoW it is a mathematical fact that the cost of an attack increases exponentially with the number of
blocks to be changed, and this cost is bounded to a physically enormous amount of energy,
the same cost for an attacker of a PoS system is unpredictable, because the cost of the ``value-at-stake''
for an attacker is not bounded to any external factor.

This position paper concludes that a \ac{PoW}-based system is more stable and predictable than a \ac{PoS}-based one,
and therefore suggests that a platform that candidates itself to be the most secure one must be \ac{PoW}-based.

\section{Do you need a Blockchain?\\$\quad$Avoid Common Pitfalls!}
\label{sec:pitfalls}

\begin{figure}
  \centering
    \includegraphics[width=\linewidth]{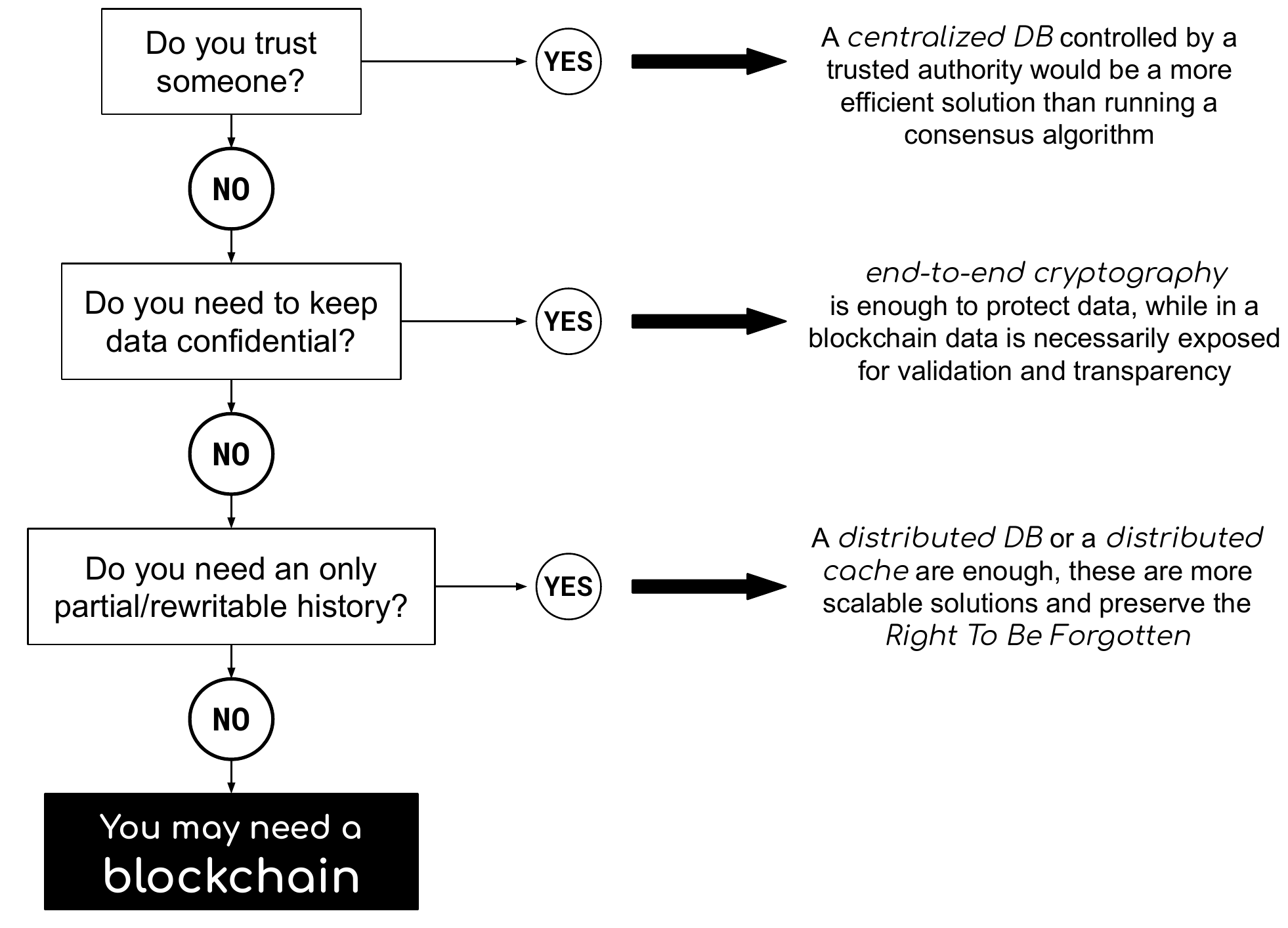} 
    \caption{Application requirements and ledger technology: Aid to decision.}
    \label{fig:whenBlockchain}
\end{figure}

A reader that accepts the blockchain defined as the open, verifiable, and immutable Shared
Ledger technology par excellence, immutable by reason of a powerful consensus protocol,
should also acknowledges it as extremely inefficient~\cite{fairley2017ridiculous,Li2019}.
For this reason, we recommend to use the blockchain technology only when needed, opting for a different 
technology whenever possible, especially for the IoT.
For example, a traditional ledger is preferable when the access is restricted to registered users,
or when data must be kept confidential, or when strong trust assumptions are given, 
which makes the strong consensus required by the blockchain an overkill.
The above considerations are illustrated in \cref{fig:whenBlockchain},
which extends the tradition started by Peck and W\"ust~\cite{peck2017,wust2018you} to provide
a guided path to clearly recognize when a blockchain is not the right choice for an application.
The one provided here distinguishes itself from previous ones~\cite{flowChartsBlockChain}
for its limited scope, i.e., highlighting the cases when the blockchain is \emph{not needed}.

\cref{fig:whenBlockchain} also highlights, implicitly, those recurrent
abuses of the blockchain in applications whose requirements conflict with the essential blockchain characteristics.
In the rest of this section we list these recurrent pitfalls to substantiate our critical analysis the blockchain applications.

\begin{Pitf}
\label{pitf:trustAssumption}
\textit{Using the blockchain in presence of strong assumptions on trust.}

Morgen E.\ Peck first identified this pitfall by analyzing the application of the blockchain to voting~\cite{peck2017},
nonetheless, many are the works that propose blockchain-based voting platforms~\cite{bcEvoting,ayed2017,bistarelli2017,liu2017voting,hardwick2018voting,wang2018large,qi2017,noizat2015}.
We note that a blockchain as defined by \cref{BCdef} contrasts the needs of voting because,
although it is true that the voter's identity should be kept secret, still users cannot be anonymous.
Their identity must be in fact uniquely determined to ensure uniform eligibility, namely, nobody should be able to cast multiple votes
(Sybil attack), \emph{``hence an identity provider is required one way or another''}~\cite{heiberg2018}.
This identity provider must be necessarily a centralized accredited institution unless, ad absurdum,
someone accept voters providing IDs issued by unofficial distributed agencies (fake documents!).
Double-voting is thus different from double-spending and is a problem whose solution imposes the existence of a
central authority: this last cannot be decentralized not even advocating a blockchain.

In general, the existence of a trusted third party (such as an identity provider)
or the assumption of mutual trust in a distributed network are considered strong assumptions.
If a trusted third party exists, then this party can play the role of the central coordinator, and by hypothesis
users can rely on it to keep a consistent \ac{DB} without going through the overhead generated by a consensus algorithm.
From the perspective of the blockchain trilemma, the need of a trusted authority is equivalent
to trade decentralization for scalability.
Clearly, under this assumption the sacrifices that come with the blockchain are not necessary.
Similarly, assuming mutual trust among nodes, there is no need to trade performance for security.
\end{Pitf}

\begin{Pitf}
\emph{Proposing a fast blockchain.} 
Fast blockchains of all kinds have been proposed, especially for business applications, circumventing the limits of the \ac{PoW} by promoting alternative protocols. 
Some popular platforms advertised as secure and fast include the HyperLedger Fabric, HyperLedger Sawtooth, Ripple, Amazon Managed Blockchain and Azure Blockchain, followed by many others. 
All these systems are collectively called Blockchain-as-a-Service (BaaS) Platforms~\cite{onik2019performance}.
However, all of them inevitably sacrifice some specific features of the blockchain in an effort to remove those that are unsuitable for the enterprise.
We raise the concern that, according to the blockchain trilemma, the enhanced performance~\cite{HypL20kTPS} 
granted by these platforms can be achieved only reducing, partially, either the decentralization or the security
of what this paper defined to be a blockchain.
Two exemplary fast platforms will be now studied to highlight the great differences with classic blockchains such as Bitcoin and \Eth.
We stress on these differences to justify, once more, the restrictive \cref{BCdef} proposed by this paper. 
Not calling ``blockchain'' these fast ledgers is a terminological choice only but we not discourage their use, rather,
we promote their adoption when a blockchain is not needed, especially in the IoT (see \cref{fig:whenBlockchain} again).

The official documentation of \textbf{HyperLedger Fabric}~\cite{hyperLdoc} describes its core design.
It also highlights that the essential characteristics of public permissionless blockchains, such as being \emph{public networks, open to anyone, where participants interact anonymously} are problematic for enterprises, especially because \emph{the identity of the participants is a hard requirement} for enterprises to comply with legal obligations such as Know-Your-Customer (KYC) and Anti-Money Laundering (AML) financial regulations.
This leads to the identification of a list of requirements for enterprises:
\begin{itemize}[itemsep=0em]
  \item Participants must be identified/identifiable.
  \item Networks need to be permissioned.
  \item High transaction throughput performance.
  \item Low latency of transaction confirmation.
  \item Privacy and confidentiality of transactions and data pertaining to business transactions.
\end{itemize}
As the HyperLedger Fabric \emph{``has been designed for enterprise use from the outset''} these requirements are all mandatory.
The following short analysis of HyperLedger Fabric (Fabric, for short) explains how it supports the listed features.

A membership service associates entities in the network with cryptographic identities.
Fabric enables Privacy and Confidentiality through its Channels architecture, where Channels are defined as \emph{private “subnet” of communication between two or more specific network members, for the purpose of conducting private and confidential transactions}~\cite{hyperChannel},
and through \textit{private data}~\cite{hyperPrivateData}.
The Fabric documentation also reports this noteworthy consideration:
\emph{By relying on the identities of the participants, a permissioned blockchain can use more traditional crash fault tolerant (CFT) or byzantine fault tolerant (BFT) consensus protocols that do not require costly mining.}
The consensus protocols supported by Fabric are indeed traditional consensus protocols of this kind.
In particular, the leader-based voting consensus protocol \textit{Raft}~\cite{ongaro2014search} is the only
non-deprecated protocol for deployments of Fabric.
With these features, that set Fabric apart from the blockchain, Fabric can reach 20\,kTPS~\cite{HypL20kTPS}, outperforming known blockchains by 4 orders of magnitude.

Observing Fabric we notice the use of more efficient and traditional consensus mechanisms,
resulting in greater performance compared to classic blockchains. 
Moreover, its security derives from a traditional access control (permissions) but not from a strong consensus mechanisms.
The resulting ledger is not open and is vulnerable to collusive attacks lead by a majority of users
not defused by any deterrent cost such as the CPU energy implied by the PoW.
As such, Fabric is more similar to traditional ledgers than to classic blockchain systems like Bitcoin.
It is in fact a decentralized platform customized for the enterprise.
That is why we include it in the class of more traditional Shared Ledgers~\cite{WhyIBMs99}. 

\textbf{HyperLedger Sawtooth}~\cite{sawtooth} is an extension of HyperLedger that explicitly requires the Intel SGX framework.
Sawtooth promotes the Proof of Elapsed Time (PoET) as a \emph{solution to the Byzantine Generals Problem that utilizes a ``trusted execution environment'' to improve on the efficiency of present solutions such as Proof-of-Work \ldots} and \emph{assumes the use of Intel SGX as the trusted execution environment}~\cite{poETdoc}.
The idea behind PoET is the following. 
A random waiting time is distributed to all nodes competing to become the next block miner.
When the waiting time expires, the node proves that it waited by providing the PoET generated by its Intel chip.
The first node that exhibits a valid PoET is elected as block-miner. 
This protocol is much more energy efficient since the Intel chip consumes much less than a Bitcoin miner to generate a PoET.
However, in this protocol users must trust the server distributing random waiting times and must also trust the proprietary Intel SGX technology, (which has been already attacked successfully multiple times because of its internal architectural flaws~\cite{sgxAttacks}). 
Moreover, if SGX chips are reasonably affordable it becomes possible for anyone to get many of them that,
together, will consume a lot of power for increasing the chances to be elected as block-miner, falling back to
a Bitcoin scenario.

In general, all \textit{fast} or \textit{low-energy} proofs facilitate attacks, so that mining-difficulty must be artificially kept high (as described for Bitcoin in \cref{subsec:freqDifficulty}) to ensure high level of security.
The trilemma warns us to beware of fast consensus protocols. They should only be run in centralized platforms, 
hence in private, non transparent organizations. Otherwise, they are prone to be easily attacked.
Therefore, applications implemented on top of BaaS platforms cannot be as secure and transparent as if implemented on top of what we defined to be a blockchain.
\end{Pitf}

\begin{Pitf}
\emph{Validating sensory data through a blockchain.}

This is a common pitfall besetting, for example, all those who choose the blockchain for supply chain management, a quintessential type of IoT application relying on sensor readings and other ``things'' from the physical world~\cite{montecchi2019s,tse2017,tian2016agri,miller2018}. 
One is lured into using the blockchain for it makes handovers among intermediate dealers manifest, from the producer to the final customer.
However, it cannot ensure that the traded goods have been transported correctly. 
As observed by W{\"u}st et al., the problem with the supply chain lies in the \emph{trust of sensors} or, in other terms, in the way trusted information is acquired from the real world~\cite{wust2018you}. 
For example, a malicious truck company that wants to cut costs of transport of refrigerated food can claim its trustworthiness showing compliance with the laws by installing ``trusted'' thermometers with GPS.
The company can then cheat installing the thermometer in a little empty fridge traveling on the truck
together with the rest of the (not refrigerated) load.
In general, the ``perception layer'' of the IoT is the most vulnerable to attacks~\cite{perceptionPorblem,lee2017blockchain}, so that IoT devices (potentially deployed outdoor without supervision) must implement in-hardware mechanisms to be secured.
Still no blockchain will ever be able to prevent all possible physical sabotages of sensors. 

We claim that sensors cannot be trusted not just because they can be compromised, but also because of the inescapable \textit{uncertainty of measurements}, independently on the source---whether benign or malicious---of the uncertainty.
Consider, for example, a smart contract used to buy train tickets that embeds in its business logic the automatic refund for travelers in the case of train delays above 30 minutes. 
It is not hard to imagine that the rail company may cheat by reporting (false) delays that are lower than 30 minutes to avoid paying refunds.
All applications that rely on measurements, taken by any kind of sensors or IoT device, at some stage must trust either the centralized company controlling that sensor or a consortium that collected the measurement.
When this happens we talk about \emph{oracles}~\cite{mostefaoui2002introduction} that must be introduced to obviate the trust problem on sensor by providing trusted information services~\cite{adler2018astraea}. 
However, these oracles are either centralized trusted authorities (see Pitfall~\ref{pitf:trustAssumption}), or systems that require distributed consensus, which reintroduce all the issues and trade-offs discussed in \cref{subsec:trilemma}.
With oracles the blockchain looses its meaning as a tool to implement distributed trust.
\end{Pitf}

\begin{Pitf}
\emph{Proposing a blockchain-based approach for confidentiality.}

Confidentiality, namely, providing data secrecy, is critical for many applications. 
There is no reason to publish and record confidential data on a public blockchain as confidentiality
is in clear contrast with a key characteristic of blockchains: Transparency, which
is given by the sum of the blockchain openness and completeness.
Works that advocate the use of the blockchain for keeping user data confidential include~\cite{dorri2017blockchain,ouaddah2017towards,rahulamathavan2017privacy,dorri2017lsb,dorri2017towards,dorri2017automotiveprivacy,dwivedi2019}. %
We note that since confidentiality contrasts with the public nature of blockchains, a careful justification of design choices is necessary.
\end{Pitf}

\begin{Pitf}
\emph{Storing sensitive information on the blockchain.}

Registering user credentials and account information on a blockchain is an irreversible operation, as data cannot be deleted.
Services implemented on top of a blockchain will not be able to delete user data; not even upon legitimate request.
This could be an issue for a user that wants to abandon a service and also for authorities that need to enforce the ``Right to be Forgotten,'' which is a legal provision in several countries~\cite{ateniese2017redactable,politou2019blockchain,berberich2016blockchain,gabison2016policy}.
\end{Pitf}

\begin{Pitf}
\emph{Verifying the authenticity of digital documents or real goods with a blockchain.}

Many proposals concern the use of the blockchain as a decentralized platform to store digitally signed documents, i.e., certificates~\cite{chang2017blockchain,grather2018blockchain,fowler2017linking,burstall2017blockchain,cheng2018blockchain}.
One might be drawn to believe that, as it is part of a blockchain, that certificate is authentic.
However, the authenticity of a certificate is guaranteed by its digital signature, which depends on a trusted authority external to the blockchain and not subject to the trust obtained via a consensus protocol.
For example, let us consider the following certificate digitally signed by an institution~$I$ and recorded in a blockchain: ``Alice passed exam~$E$ after attending the course provided by Institution~$I$ on date~$D$.'' 
The fact that this certificate belongs to a blockchain does not in any way prove that Alice really attended a course and passed an exam, and legitimate doubts can also arise about the timestamped date~$D$. 
Trusting the authenticity of this document only relies on unconditionally trusting the issuer, namely, institution~$I$.
Overall, the blockchain can highlight how a series of operations is chronologically consistent, thus valid.
However, for non transactional data like common certificates, patents and proofs of authorship, the blockchain cannot
assist their authentication or validation.

This holds also for identity documents, as argued in \cref{pitf:trustAssumption}, and for real goods too.
Somehow, in fact, many have been inspired by the blockchain immutability to guarantee the authenticity of real goods~\cite{montecchi2019s,tse2017,tian2016agri,nowinski2017can,galvez2018future,bahga2016blockchain,boehm2017holistic}.
The idea would be to associate any good with a digital identifier (e.g., a QR code) tracked by a blockchain so that a final customer receiving the good can use the attached identifier as a proof of authenticity.
Unluckily, there is no mechanical tool to make objects of the real world unforgeable: For as much as the identifier on the blockchain is immutable (unforgeable), this is not true for the associated good, which can be physically replaced with another of lower quality.
We stress that the \emph{digital signature} is the technology for verifying the authenticity of transactions or digital documents
but the blockchain alone, instead, cannot prove the authenticity of products or certificates.
\end{Pitf}

\begin{Pitf}
\emph{Forgetting the cost of mining.}
One can think to the blockchain as the remover of banking fees.
This is true as long as an incentive mechanism encourage validators in processing transactions, 
but turns to false as soon as this incentive mechanism terminates.
Any blockchain project should describe a sound incentive mechanism for miners and consider transaction costs
or it will be destined to failure, as no actors of the system will bear the cost of mining. 
\end{Pitf}

\begin{Pitf}
\emph{Implementing a memoryless process on top of a blockchain.}

In a typical blockchain system the full chain of transactions must be recorded for validating new transactions. 
For those applications designed on top of Markovian assumptions, i.e., where only the knowledge of the \textit{current state} of the system is necessary to make progress, using a blockchain will keep recording (possibly too many) unnecessary data.
For example, there is no need for a smart home IoT application to keep memory of the full history of the temperature of a room to decide which heat source to activate or disable at the current time.
Even accounting and billing does not require the entire history, but only a previous reading and recent invoicing. 
\end{Pitf}

\begin{Pitf}
\emph{Claiming to jointly provide maximum security, decentralization and performance.}

Given the current state of science and technology, the proposals claiming the joint provision of maximum
security, decentralization and performance violate the limits of distributed systems (\cref{subsec:limitsConsensus}).
By blockchain trilemma they cannot truly work as promised. 
This is witnessed by the fact that, among the blockchain projects launched in the past few years, many have already failed~\cite{deadCryptocurrencies} and some have been even denounced as Ponzi schemes~\cite{bartoletti2020dissecting}.
\end{Pitf}

\section{Blockchain in Support of the IoT}
\label{sec:roleBCwithIoT}

\begin{figure}[]
  \centering
    \includegraphics[width=\linewidth]{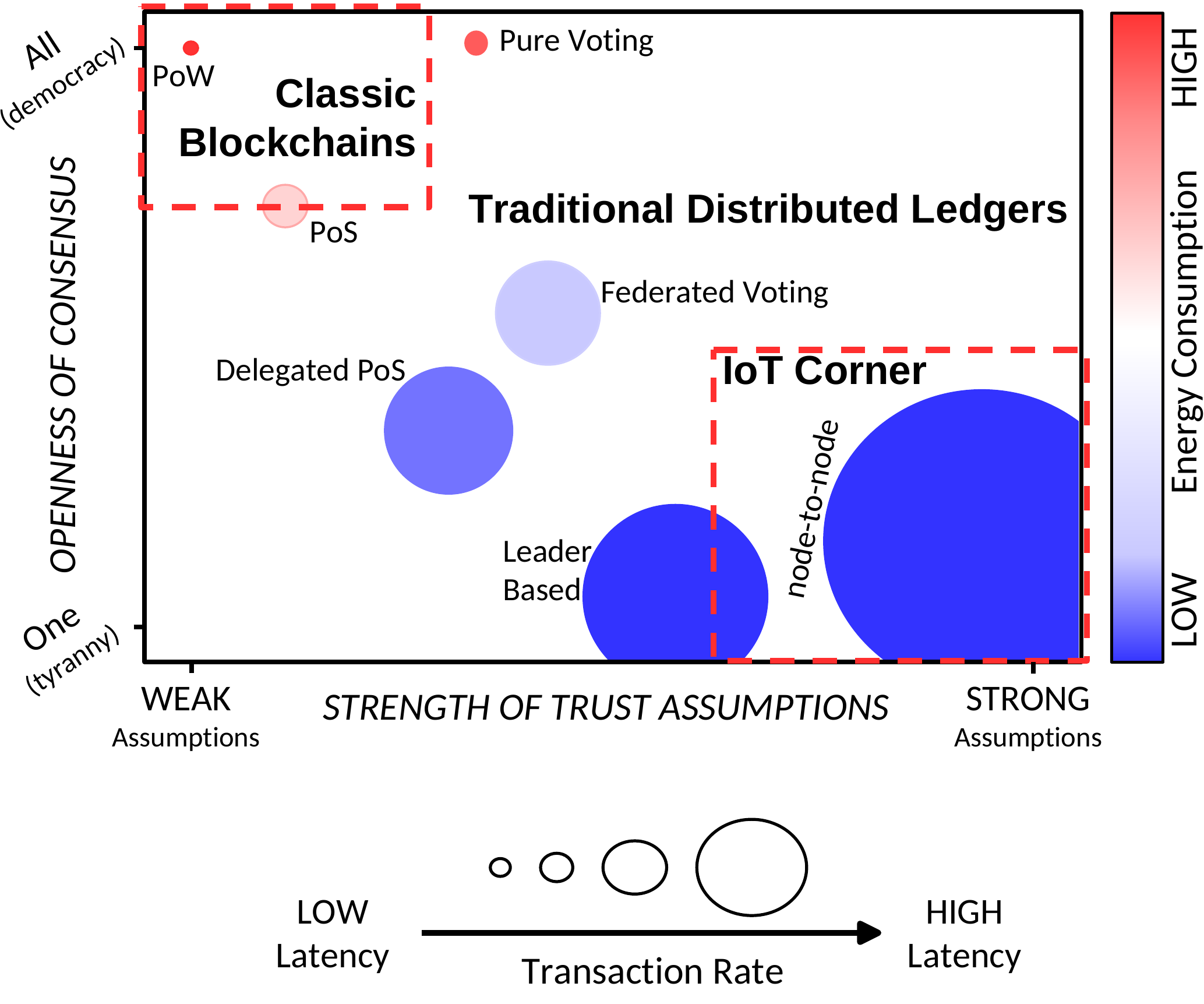} 
    \caption{Bubblechart of relevant building blocks for Shared Ledgers. The diagram compares solutions as a function of trustworthiness of users (x-axis) and openness of the consensus protocol (y-axis). The color and the size of each bubble offer a quick indication of energy consumption and transaction rate of each consensus protocol, respectively.}
    \label{fig:bubbleChart}
\end{figure}

Our exploration of the trade offs imposed by the blockchain trilemma (\cref{fig:trilemma}) together with the 
clarification of the applications that, in light of our blockchain definition, result to be flawed,
do not hint to any promising strategy for the \textit{integration} of the blockchain within the IoT.
However, we argue that the blockchain can still be used as an \emph{external service} supporting the decentralized validation
of IoT transactions, offering a complementary or alternative paradigm to centralized cloud services. 

To illustrate how the blockchain can successfully play as an external ---not integrated--- ledger for the IoT,
we introduce the ``bubblechart'' of \cref{fig:bubbleChart}, which draws a multidimensional overview of the 
consensus mechanisms surveyed in \cref{subsec:shortRevConsensus} according to the following four dimensions:

\begin{itemize}[itemsep=0em, leftmargin=0.5cm]

\item The strength of the trust assumptions, which is inversely proportional to the degree of security (x-axis).

\item The openness of consensus, an indicator of the degree of democracy, from one (tyranny) to all (power to the people) (y-axis).

\item Energy consumption (color of each bubble: Red when high or blue if low).

\item The transaction rate (size of each bubble: The larger the faster). 
The transaction rate can be considered also an indicator of transaction latency: Fixing the number of transactions registered with a single operation (block-size), a lower latency leads to a higher transaction rate.

\end{itemize}

The figure enriches the trilemma by breaking down the ``scalability vertex'' (\cref{fig:trilemma}) into two distinct dimensions, i.e., energy consumption and transaction rate. 

The ideal ``blockchain-for-IoT'' bubble would be blue and large (low-power and very performing) in the top-left
(most decentralized and most secure) corner of the chart.
Blockchain systems are naturally located in the top-left corner, characterized by being democratically open and secure despite weak trust assumptions, but also extremely slow and resource-hungry. 
The opposite corner is where IoT applications reside, with their scalability requirements, tight resource constraints, high global transaction rates.
This corner also highlights that the ultimate participants are ``things'' rather than humans, thus affording a higher degree of trust, at least towards some other parts of the system.
As such, \Cref{fig:bubbleChart} pictorially conveys our crucial observation: Classic blockchains are in contrast with IoT requirements, 
which keeps the two realms well separated.

However, despite the lack of space for blockchain systems in the bottom-right corner of our bubblechart, going back to \cref{tab:consensusComparison} (last line), we identify \emph{node-to-node consensus} as a means to build trust among operators/systems without established relations.
The key word, here, is \textit{node-to-node}, which is used to restrict the distributed consensus problem to few nodes, usually a couple although extensions to small numbers is efficiently conceivable.
To settle a transaction it is sufficient for the transacting parties to agree on the transaction protocol, and this agreement can be reached privately by the two (or few more) parties in any fashion.
What makes node-to-node consensus appealing for IoT is its efficient support of \textit{local consensus}, which is natural for many IoT applications such as those with groups of sensors or a platoon of vehicles.

The number of different node-to-node consensus protocols is limited only by imagination.
Specific transitive properties, i.e., how and to what extent if node~A trusts node~B and node~B trusts node~C, then node~A can trust node~C,
can be defined to be applied to large clusters of trusted entities, ultimately leading to a network (the IoT itself in some sense) of diverse but interoperating ``channels.''
We inherit the term ``channel'' from the world of cryptocurrencies (\textit{Networks of Payment Channels}~\cite{gscn-ccs18,towardsPayNTW,bcIoTpaymensSurv,fasterPayNTW,PaymentChannel}) and from that of communication networks, where a network is a set of channels interconnecting its nodes. 
Since IoT transactions are not necessarily monetary, in the rest of this work we use the generic term \textit{Transaction Channels}. 
In the remainder of this section we describe the networks of Transaction Channels, and how they enable the external use of the blockchain for the IoT.

\subsection{\textbf{Networks of Transaction Channels}} 

\textit{Transaction Channels} are all those techniques used to group off-chain transactions between the same small group of users to speed them up and avoid paying multiple transaction fees. 
A Transaction Channel is therefore a node-to-node consensus protocol where the two transacting parties establish a fast ``payment'' method and agree to postpone the clearing of the transactions' balance.
Recording the status of the channel on the blockchain can be periodic or event-based.
What is stored in the blockchain is a meaningful representation of the transactions' history. 
For example, this could be the stochastic representation of a long-term distributed measure or the amount of energy exchanged in a smart grid.

The most notable implementation of Transaction Channels is the Lightning Network~\cite{poon2016bitcoin}, which scales up the technique to a full network of such channels. 
This allow payments to be routed between remote end-points thanks to a dedicated routing protocol for payments, with the aim of providing a globally scalable mechanism for fast transactions. 
The Lightning Network is ``Bitcoin oriented,'' but the concept of a network of payment channels may become the transaction platform enabling a global market at the IoT scale.
It also opens the way to thrilling research challenges such as bringing network science and expertise into the domain of transporting and routing payments within Payment Networks, as explored in~\cite{sivaraman2018high}.
Major open problems include addressing the depletion of channel capacity, especially for the most loaded nodes in the center of the network, developing enhanced centrality-aware routing strategies~\cite{MaLo18Pop,maccari2020exact}, and rebalancing techniques~\cite{pickhardt2019imbalance,tikhomirov2020probing,eckeysplitting}.

\subsection{The Role of the Blockchain in the IoT}

\begin{figure}[h]
  \centering
    \includegraphics[scale=.4]{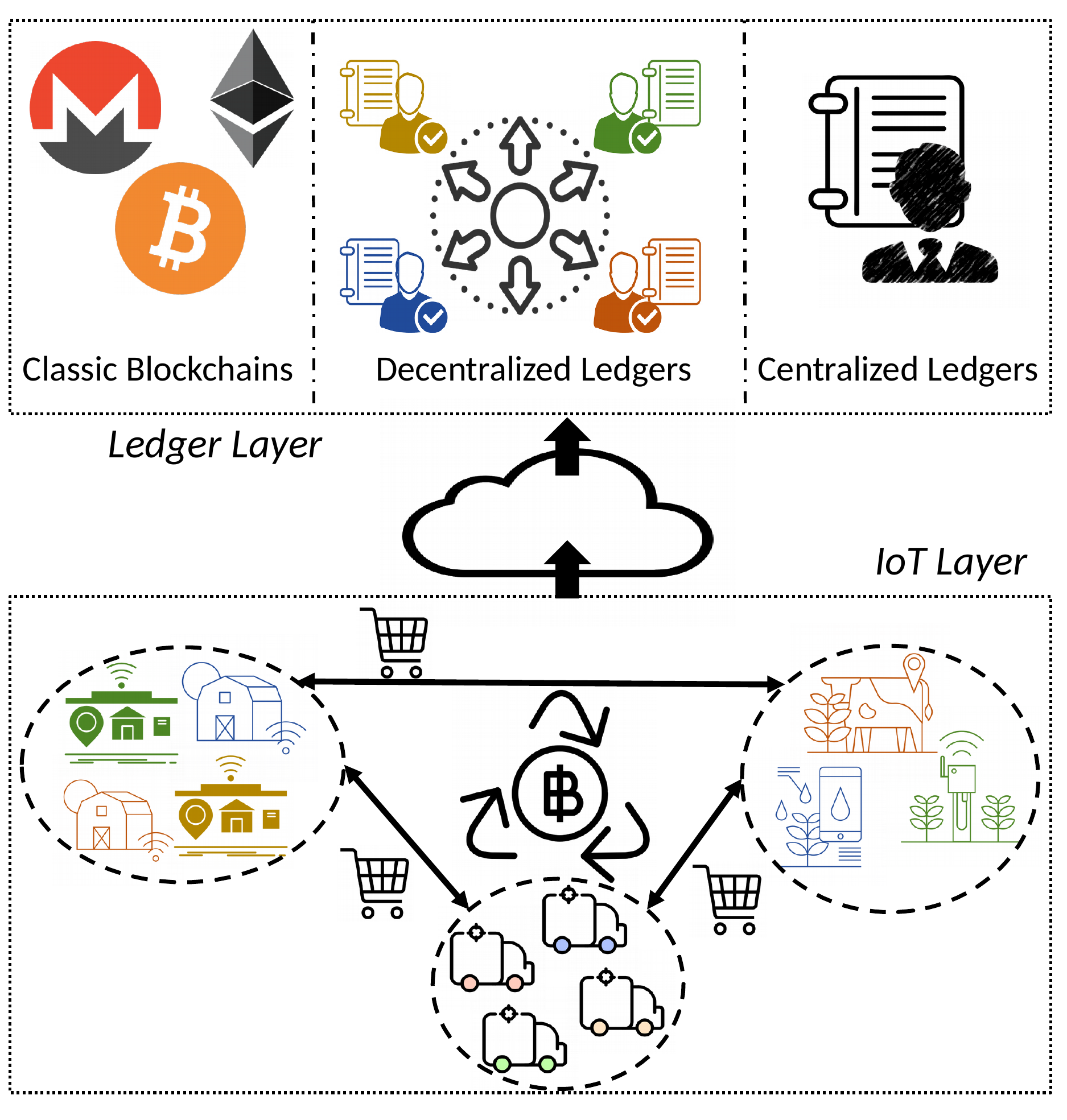} 
    \caption{Different IoT clusters, made of devices managed either by private (home/enterprises) or public (institutional) entities perform most transactions locally, in the IoT layer. For interoperability purposes the different IoT clusters access intermediary platforms. 
    Blockchain-based platforms can provide specific services at this level in line with their characteristic capabilities of granting security and immutability even in the absence of trust.} 
    \label{fig:bcIoT}
\end{figure}
 
\cref{fig:bcIoT} illustrates our vision of the IoT empowered by Networks of Transaction Channels deployed at the IoT layer.
Here blockchains can play as supporting external ledgers, similarly to how the Bitcoin blockchain supports
the recording of the channels status in the Lightning Network.

This vision stems from the observation that most IoT applications will have a \textit{local} relevance.
For example, domotic devices will intercommunicate mostly only over a house local network.
Similarly, agricultural sensors for precision farming will mostly communicate with each other and convey the local information to a gateway controlled by the farmer. 
Industrial IoT often requires high levels of privacy and confidentiality, clearly in contrast with the open, immutable nature of a blockchain; vehicular networks and intelligent transportation systems may require transactions with latency smaller than a few milliseconds, and rates in the order of kTPS per vehicle, again in full contrast with the characteristics of blockchains.
IoT transactions are local and normally lightweight in nature, therefore calling for local and lightweight solutions for the platform to support them. 
From time to time, separate IoT domains, platforms and applications may need to carry out and record transactions with a global, final, and immutable nature. 
At this level blockchains will play an important role, freeing IoT systems from the need to subscribe to a global, centralized, expensive, trust-based service whose security and reliability have well-known limitations.
Using blockchains externally would therefore bring added value to the IoT domain, responding to its requirements of extending beyond local, context-limited applications when needed.

\section{Conclusions}
\label{sec:conclusion}

In this paper we argue that the blockchain is not the appropriate technology for securing the IoT, albeit it can bring added value as an external service.
To support this claim we made some clarity around the very name ``blockchain,'' to dispel the many misunderstandings that hamper its usage and makes it appear as a universal---almost magic---technology.

For this purpose we explore consensus protocols, reviving those theorems that give rise to fundamental trade-offs such as the emblematic blockchain trilemma.
We raise the concern that stake-based protocols fully rely on the rationality assumption of their users and, compared to traditional voting based protocols, lack of mathematical stability properties.
This means that stake-based systems are prone to market failures and bubbles like real stock markets---a very dangerous risk. 
Moreover, we stress how the voting power has a tendency to consolidate in the hands of few great stakeholders with both PoS and PoW.
For this reason, we suggest to consider the two as \textit{census suffrage} mechanisms, with PoS preferable over PoW to reduce
the energy consumption. 
However, the immutability of a PoS-blockchain is questionable, so a PoW-based one is considered more secure.
In the landscape of the Shared Ledger technologies that we draw from the distributed system perspective of the IoT, we highlight the innovative and peculiar aspects of permissionless blockchains in contrast with permissioned ones, the latter turning out to be not so different from traditionally managed data bases.
We conclude that the term ``blockchain'' should be reserved to those platforms characterized by:
\textit{i)} openness to anonymous users;
\textit{ii)} full and public history of transactions, and 
\textit{iii)} safeguarded by a strong consensus protocol.
This definition has far-reaching consequences. 
Above all, the strong consensus protocol requirement necessarily brings high resource consumption to counter the lack of trust between users, and imposes transactions rates and latency unacceptable for most IoT scenarios.
As such, we stress that the blockchain is not a technology suitable for popular applications such as e-voting and supply chain management.
Furthermore, to not violate the Right to Be Forgotten, the blockchain can hardly support Identity Management applications
nor work as archive of certificates and other sensitive information. 

In conclusion, we advocate using the blockchain only in those IoT scenarios where the transactions are supported by local, lightweight platforms whose consensus is tailored to the domain of application and the local context.
We name these platforms ``Transaction Channels.''
These channels may (or may not, depending on the application) interact through aggregate, rare transactions to form a global network of Transaction Channels, which can be successfully based on the blockchain technology, freeing the IoT from the need to rely on global, centralized platforms to interact across diverse application, technology, and administrative domains. 

\setlength{\bibsep}{0pt plus 0.3ex}
\footnotesize
%https://github.com/ionaic/bibtex-ieeetran-urldate
\bibliographystyle{IEEEtranUrldate.bst}
\bibliography{ms.bib}
\end{document}